\documentclass[reprint,amsmath,amssymb,aps,prd,nofootinbib]{revtex4-1}

\usepackage{graphicx}
\usepackage{color}
\usepackage{dcolumn}
\usepackage{bm}
\usepackage{hyperref}
\usepackage{paralist}
\usepackage{amsmath}
\usepackage{bm}

\def\bra#1{\mathinner{\langle{#1}|}}
\def\ket#1{\mathinner{|{#1}\rangle}}

\begin{document}

\preprint{APS/123-QED}

\title{Quantum field theory of relic nonequilibrium systems}

\author{Nicolas G. Underwood}
\email{nunderw@clemson.edu}
\author{Antony Valentini}%
 \email{antonyv@clemson.edu}
\affiliation{Department of Physics and Astronomy,\\
Clemson University, Kinard Laboratory,\\
Clemson, SC 29634-0978, USA}

\begin{abstract}
In terms of the de Broglie-Bohm pilot-wave formulation of quantum theory, we develop field-theoretical models of quantum nonequilibrium systems which could exist today as relics from the very early universe. We consider relic excited states generated by inflaton decay, as well as relic vacuum modes, for particle species that decoupled close to the Planck temperature. Simple estimates suggest that, at least in principle, quantum nonequilibrium could survive to the present day for some relic systems. The main focus of this paper is to describe the behaviour of such systems in terms of field theory, with the aim of understanding how relic quantum nonequilibrium might manifest experimentally. We show by explicit calculation that simple perturbative couplings will transfer quantum nonequilibrium from one field to another (for example from the inflaton field to its decay products). We also show that fields in a state of quantum nonequilibrium will generate anomalous spectra for standard energy measurements. Possible connections to current astrophysical observations are briefly addressed.
\end{abstract}

\maketitle


\section{Introduction}
\label{I}
In the de Broglie-Bohm pilot-wave formulation of quantum theory
\cite{deB28,BV09,B52a,B52b,Holl93}, the Born probability rule has a dynamical
origin \cite{AV91a,AV92,AV01,VW05,EC06,TRV12,SC12} and ordinary quantum
physics is recovered as a special equilibrium case of a wider nonequilibrium
physics
\cite{AV91a,AV91b,AV92,AV96,AV01,AV02,AV07,AV08,AV09,AV10,AVPwtMw,PV06}. On
this view, we may understand the Born rule as arising from a relaxation
process that took place in the remote past. Quantum nonequilibrium -- that is,
violations of the Born rule -- may have existed in the very early universe
before relaxation took place \cite{AV91a,AV91b,AV92,AV96}. Such effects could
leave observable traces today -- in the cosmic microwave background (CMB)
\cite{AV07,AV08,AV09,AV10,CV13,CV14} or in relic systems that decoupled at
very early times \cite{AV01,AV07,AV08}. The former possibility has been
developed in some detail and comparisons with data are beginning to be made
\cite{AV10,CV13,CV14,PVV14}. The latter possibility is the focus of this paper.

According to our current understanding, the observed temperature anisotropy in
the CMB was ultimately seeded by quantum fluctuations during an inflationary
era \cite{LL00,Muk05,W08,PU09}. Inflationary cosmology then provides us with
an empirical window onto quantum probabilities in the very early universe. On
an expanding radiation-dominated background, relaxation in pilot-wave theory
can be suppressed at long (super-Hubble)\ wavelengths while proceeding
efficiently at short (sub-Hubble)\ wavelengths
\cite{AV07,AV08,AV10,CV13,CV14,AVbook}. Thus, in a cosmology with a
radiation-dominated pre-inflationary phase \cite{VF82,L82,S82,PK07,WN08}, one
may obtain a large-scale or long-wavelength power deficit in the CMB
\cite{AV07,AV08,AV10,CV13,CV14}. For an appropriate choice of cosmological
parameters, the expected deficit is consistent with the deficit found in data
from the \textit{Planck} satellite \cite{PlanckXV,CV13,CV14}. Whether the
observed deficit is in fact caused by quantum relaxation suppression during a
pre-inflationary era or by some other more conventional effect remains to be seen.

A pilot-wave or de Broglie-Bohm treatment of the early Bunch-Davies vacuum
shows that relaxation to quantum equilibrium does not take place at all during
inflation itself \cite{AV07,AV10}. Thus, if a residual nonequilibrium still
existed at the end of a pre-inflationary era, the inflaton field would carry
traces of that nonequilibrium forward to much later times.\textbf{ }Similarly,
should nonequilibrium be generated during the inflationary era by exotic
gravitational effects at the Planck scale \cite{AV10}, the resulting
departures from the Born rule will be preserved in the inflaton field and
carried forward into the future where they might have an observable effect.

As we shall discuss in this paper, as well as imprinting a power deficit onto
the CMB sky, a nonequilibrium inflaton field would also transfer
nonequilibrium to the particles that are created by inflaton decay. Since such
particles make up almost all of the matter present in our universe, it seems
conceivable that today there could exist relic particles that are still in a
state of quantum nonequilibrium. We will also consider relic vacuum modes for
other fields (apart from the inflaton) as potential carriers of nonequilibrium
at late times.

These scenarios raise a number of immediate questions. First of all, even if
nonequilibrium relics were created in the early universe, could the
nonequilibrium survive until late times and be detected today? As we shall
see, simple estimates suggest that (at least in principle) relaxation to
equilibrium could be avoided for some relic systems. A second question that
must be addressed is the demonstration, in pilot-wave field theory, that
perturbative interactions will in general transfer nonequilibrium from one
field to another. This will be shown for a simplified model of quantum field
theory involving just two energy levels for each field. Finally, one must ask what kind of
new phenomena might be generated by relic nonequilibrium systems in an
astrophysical or cosmological context. This opens up a potentially large
domain of investigation. General arguments have already shown that the
quantum-theoretical predictions for single-particle polarisation probabilities
(specifically Malus' law) would be broken for nonequilibrium systems
\cite{AV04a}. In this paper we focus on measurements of energy as a simplified
model of high-energy processes. It will be shown that conventional energy
measurements performed on nonequilibrium systems would generate anomalous
 spectra. We may take this as a broad indication of the kinds of
anomalies that would be seen in particle-physics processes taking place in the
presence of quantum nonequilibrium.

In this paper we are not concerned with the question of practical detection of
relic nonequilibrium. Rather, our intention is to make a case that detection
might be possible at least in principle, and to begin the development of
field-theoretical models of the behaviour of relic nonequilibrium matter.

Generally speaking, even a lowest-order calculation of perturbative processes
in quantum field theory will involve all of the field modes that are present
in the system. While such calculations are in principle possible in de
Broglie-Bohm theory, in practice it would involve integrating trajectories for
an unlimited number of field modes. In this paper, we make a beginning by
confining ourselves to simplified or truncated models of quantum field theory
involving only a small number of field modes. Our models are inspired by
approximations commonly used in quantum optics, where one is often interested
in the dynamics of a single (quantised)\ electromagnetic field mode inside a
cavity. Our main aim is to justify the assertions that underpin our scenarios.
In particular, we wish to show by explicit calculation of examples that
perturbative couplings will in general transfer nonequilibrium from one field
to another, and that nonequilibrium will affect the spectra for
basic particle-physics processes involving measurements of energy. We
emphasise that the calculations presented in this paper are only intended to
be broadly illustrative. The development of more realistic models is left for
future work.

In Section \ref{IIA} we summarise the background to our scenario, and in particular
the justification for why the inflaton field singles itself out as a natural
carrier of primordial quantum nonequilibrium. In Section \ref{IIB} we argue that
inflaton decay can generate particles in a state of quantum nonequilibrium
(induced by nonequilibrium inflaton perturbations and also by the other
nonequilibrium degrees of freedom that can exist in the vacuum), and that such
nonequilibrium could in principle survive to the present day for those decay
particles that were created at times later than the relevant decoupling time.
The gravitino provides a suggestive, or at least illustrative, candidate. In
Section \ref{IIC} we consider a somewhat simpler scenario involving relic
nonequilibrium field modes for the vacuum only. For simplicity we restrict
ourselves to conformally-coupled fields, as these will not be excited by the
spatial expansion. It is argued that super-Hubble vacuum modes that enter the
Hubble radius after the decoupling time for the corresponding particle species
will remain free of interactions and could potentially carry traces of
primordial nonequilibrium to the present day (for sufficiently long comoving
wavelengths). An illustrative example is provided by the massless gravitino.
In Section \ref{IID} we indicate how particle-physics processes would be affected by
a nonequilibrium vacuum.

These preliminary considerations provide motivation for the subsequent
detailed calculations. In Section \ref{particle_decay} we give an example of the perturbative
transfer of nonequilibrium from one field to another, a process that could
play a role in inflaton decay as well as in the decay of relic nonequilibrium
particles generally. In Section \ref{IV} we provide a field-theoretical model of
energy measurements, and we show by detailed calculation of various examples
that nonequilibrium will entail corrections to the energy spectra generated by
high-energy physics processes. Finally, in Section V we present our
conclusions. We briefly address the possible relevance of our scenarios to
current searches for dark matter. We also comment on some practical obstacles
to detecting relic nonequilibrium (even if it exists) and we emphasise the
gaps in our scenarios that need to be filled in future work.

\section{Relic nonequilibrium systems}

\label{II}
In this section we first summarise the background to our scenario and in
particular the role that quantum nonequilibrium might play in the very early
universe. We then provide some simple arguments suggesting that primordial
violations of the Born rule might survive until much later epochs and perhaps
even to the present day \cite{AVbook}. These arguments motivate the detailed
analysis of nonequilibrium systems provided later in the paper.

\subsection{Nonequilibrium primordial perturbations}
\label{IIA}

In de Broglie-Bohm pilot-wave theory \cite{deB28,BV09,B52a,B52b,Holl93}, a
system has a configuration $q(t)$ whose velocity $\dot{q}\equiv dq/dt$ is
determined by the wave function $\psi(q,t)$. As usual, $\psi$ obeys the
Schr\"{o}dinger equation $i\partial\psi/\partial t=\hat{H}\psi$ (with
$\hbar=1$). For standard Hamiltonians $\dot{q}$ is proportional to the phase
gradient $\operatorname{Im}\left(  \partial_{q}\psi/\psi\right)  $. Quite
generally,
\begin{align}\label{the_first}
\frac{dq}{dt}=\frac{j}{|\psi|^{2}}
\end{align}
where $j=j\left[  \psi\right]  =j(q,t)$ is
the Schr\"{o}dinger current \cite{SV08}. The configuration-space `pilot wave'
$\psi$ guides the motion of an individual system and has no intrinsic
connection with probabilities. For an ensemble with the same wave function we
may consider an arbitrary distribution $\rho(q,t)$ of configurations $q(t)$.
By construction, $\rho(q,t)$ will satisfy the continuity equation
\begin{align}\label{ant_cont}
\frac{\partial\rho}{\partial t}+\partial_{q}\cdot\left(  \rho\dot{q}\right)  =0.
\end{align}
Because $\left\vert \psi\right\vert ^{2}$ obeys the same equation, an initial
`quantum equilibrium' distribution $\rho(q,t_{i})=\left\vert \psi
(q,t_{i})\right\vert ^{2}$ trivially evolves into a final quantum equilibrium
distribution $\rho(q,t)=\left\vert \psi(q,t)\right\vert ^{2}$. In equilibrium
we obtain the Born rule and the usual empirical predictions of quantum theory
\cite{B52a,B52b}. Whereas, for a nonequilibrium ensemble ($\rho(q,t)\neq
\left\vert \psi(q,t)\right\vert ^{2}$), the statistical predictions will
generally differ from those of quantum theory
\cite{AV91a,AV91b,AV92,AV96,AV01,AV02,AV07,AV08,AV09,AV10,AVPwtMw,PV06}.

If they existed, nonequilibrium distributions would generate new phenomena
that lie outside the domain of conventional quantum theory. This new physics
would allow nonlocal signalling \cite{AV91b} -- which is causally consistent
with an underlying preferred foliation of spacetime \cite{AV08a} -- and it
would also allow `subquantum' measurements that violate the uncertainty
principle and other standard quantum constraints \cite{AV02,PV06}.

The equilibrium state $\rho=\left\vert \psi\right\vert ^{2}$ arises from a
dynamical process of relaxation (roughly analogous to thermal relaxation).
This may be quantified by an $H$-function $H=\int dq\ \rho\ln(\rho/\left\vert
\psi\right\vert ^{2})$ \cite{AV91a,AV92,AV01}. Extensive numerical simulations
have shown that when $\psi$ is a superposition of energy eigenstates there is
rapid relaxation $\rho\rightarrow\left\vert \psi\right\vert ^{2}$ (on a
coarse-grained level) \cite{AV92,AV01,VW05,EC06,TRV12,SC12,ACV14}, with an
approximately exponential decay of the coarse-grained $H$-function with time
\cite{VW05,TRV12,ACV14}. In this way, the Born rule arises from a relaxation
process that presumably took place in the very early universe
\cite{AV91a,AV91b,AV92,AV96}. While ordinary laboratory systems -- which have
a long and violent astrophysical history -- are expected to obey the
equilibrium Born rule to high accuracy, quantum nonequilibrium in the early
universe can leave an imprint in the CMB \cite{AV07,AV10,CV13,CV14} and
perhaps even survive in relic particles that decoupled at sufficiently early
times \cite{AV01,AV07,AV08}. The latter possibility provides the subject
matter of this paper.

Much of the physics may be illustrated by the dynamics of a massless,
minimally-coupled and real scalar field $\phi$ evolving freely on an expanding
background with line element $d\tau^{2}=dt^{2}-a^{2}d\mathbf{x}^{2}$ (where
$a=a(t)$ is the scale factor and we take $c=1$). Beginning with the classical
Lagrangian density%
\begin{align}
\mathcal{L}=\frac{1}{2}\sqrt{-g}g^{\mu\nu}\partial_{\mu}\phi\partial_{\nu}%
\phi\ ,
\end{align}
where $g_{\mu\nu}$ is the background metric, and working with Fourier
components $\phi_{\mathbf{k}}=\frac{\sqrt{V}}{(2\pi)^{3/2}}\left(
q_{\mathbf{k}1}+iq_{\mathbf{k}2}\right)  $ -- where $V$ is a normalisation
volume and $q_{\mathbf{k}r}$ ($r=1,2$) are real variables -- the field
Hamiltonian becomes a sum $H=\sum_{\mathbf{k}r}H_{\mathbf{k}r}$ where%
\begin{align}
H_{\mathbf{k}r}=\frac{1}{2a^{3}}\pi_{\mathbf{k}r}^{2}+\frac{1}{2}%
ak^{2}q_{\mathbf{k}r}^{2}%
\end{align}
is formally the Hamiltonian of a harmonic oscillator with mass $m=a^{3}$ and
angular frequency $\omega=k/a$. Straightforward quantisation then yields the
Schr\"{o}dinger equation%
\begin{align}
i\frac{\partial\Psi}{\partial t}=\sum_{\mathbf{k}r}\left(  -\frac{1}{2a^{3}%
}\frac{\partial^{2}}{\partial q_{\mathbf{k}r}^{2}}+\frac{1}{2}ak^{2}%
q_{\mathbf{k}r}^{2}\right)  \Psi
\end{align}
for the wave functional $\Psi=\Psi\lbrack q_{\mathbf{k}r},t]$, from which one
may identify the de Broglie guidance equation%
\begin{align}
\frac{dq_{\mathbf{k}r}}{dt}=\frac{1}{a^{3}}\operatorname{Im}\frac{1}{\Psi
}\frac{\partial\Psi}{\partial q_{\mathbf{k}r}}%
\end{align}
for the evolving degrees of freedom $q_{\mathbf{k}r}$ \cite{AV07,AV08,AV10}.
(We have assumed a preferred foliation of spacetime with time function $t$. A
similar construction may be given in any globally-hyperbolic spacetime
\cite{AV04b,AV08a,AVbook}.)

An unentangled mode $\mathbf{k}$ has an independent dynamics with wave
function $\psi_{\mathbf{k}}(q_{\mathbf{k}1},q_{\mathbf{k}2},t)$. The equations
are the same as those for a nonrelativistic two-dimensional harmonic
oscillator with time-dependent mass $m=a^{3}$ and time-dependent angular
frequency $\omega=k/a$. Thus we may discuss relaxation for a single field mode
in terms of relaxation for such an oscillator \cite{AV07,AV08}. It has been
shown that the time evolution is mathematically equivalent to that of a
standard oscillator (with constant mass and constant angular frequency) but
with real time $t$ replaced by a `retarded time' $t_{\mathrm{ret}}(t)$ that
depends on the wavenumber $k$ \cite{CV13}. Thus, in effect, cosmological
relaxation for a single field mode may be discussed in terms of relaxation for
a standard oscillator.

Cosmological relaxation has been studied in detail for the case of a
radiation-dominated expansion, with $a\propto t^{1/2}$ \cite{CV13,CV14}. In
the short-wavelength or sub-Hubble limit, it is found that $t_{\mathrm{ret}%
}(t)\rightarrow t$ and so we obtain the time evolution of a field mode on
Minkowski spacetime, with rapid relaxation taking place for a superposition of
excited states. On the other hand, for long (super-Hubble) wavelengths it is
found that $t_{\mathrm{ret}}(t)<<t$ and so relaxation is
retarded.\footnote{Such retardation may also be described in terms of the mean
displacement of the trajectories \cite{AV08,AVbook}.} Thus, in a cosmology
with a radiation-dominated pre-inflationary era, at the onset of inflation we
may reasonably expect to find relic nonequilibrium at sufficiently large
wavelengths \cite{AV10,CV13,CV14}.

No further relaxation takes place during inflation itself. This has been shown
by calculating the de Broglie-Bohm trajectories of the inflaton field in the
Bunch-Davies vacuum \cite{AV07,AV10}. In terms of conformal time $\eta=-1/Ha$,
the wave functional is simply a product $\Psi\lbrack q_{\mathbf{k}r}%
,\eta]=\prod\limits_{\mathbf{k}r}\psi_{\mathbf{k}r}(q_{\mathbf{k}r},\eta)$ of
contracting Gaussian packets and the trajectories take the simple form
$q_{\mathbf{k}r}(\eta)=q_{\mathbf{k}r}(0)\sqrt{1+k^{2}\eta^{2}}$. The time
evolution of an arbitrary nonequilibrium distribution $\rho_{\mathbf{k}%
r}(q_{\mathbf{k}r},\eta)$ then amounts trivially to the same overall
contraction that occurs for the equilibrium distribution. It follows that the
width of the evolving nonequilibrium distribution remains in a constant ratio
with the width of the evolving equilibrium distribution. Thus the ratio%
\begin{align}
\xi(k)\equiv\frac{\left\langle |\phi_{\mathbf{k}}|^{2}\right\rangle
}{\left\langle |\phi_{\mathbf{k}}|^{2}\right\rangle _{\mathrm{QT}}}
\label{ksi}%
\end{align}
of the nonequilibrium variance $\left\langle |\phi_{\mathbf{k}}|^{2}%
\right\rangle $ to the quantum-theoretical variance $\left\langle
|\phi_{\mathbf{k}}|^{2}\right\rangle _{\mathrm{QT}}$ is preserved in time. Any
relic nonequilibrium ($\xi\neq1$) that exists at the beginning of inflation is
preserved during the inflationary era and is simply transferred to larger
lengthscales as physical wavelengths $\lambda_{\mathrm{phys}}=a\lambda
=a(2\pi/k)$ grow with time.

It follows that incomplete relaxation at long wavelengths during a
pre-inflationary era can change the spectrum of perturbations during inflation
and thus affect the primordial power spectrum for the curvature perturbations
that seed the temperature anisotropy in the CMB. An inflaton perturbation
$\phi_{\mathbf{k}}$ generates a curvature perturbation $\mathcal{R}%
_{\mathbf{k}}\propto\phi_{\mathbf{k}}$ (where $\phi_{\mathbf{k}}$ is evaluated
at a time a few e-folds after the mode exits the Hubble radius) \cite{LL00}.
This in turn generates the observed angular power spectrum%
\begin{align}
C_{l}=\frac{1}{2\pi^{2}}\int_{0}^{\infty}\frac{dk}{k}\ \mathcal{T}%
^{2}(k,l)\mathcal{P}_{\mathcal{R}}(k) \label{Cl2}%
\end{align}
for the CMB, where $\mathcal{T}(k,l)$ is the transfer function and%
\begin{align}
\mathcal{P}_{\mathcal{R}}(k)\equiv\frac{4\pi k^{3}}{V}\left\langle \left\vert
\mathcal{R}_{\mathbf{k}}\right\vert ^{2}\right\rangle \label{PPS}%
\end{align}
is the primordial power spectrum. From (\ref{ksi}) we have%
\begin{align}
\mathcal{P}_{\mathcal{R}}(k)=\mathcal{P}_{\mathcal{R}}^{\mathrm{QT}}(k)\xi(k)
\label{xi2}%
\end{align}
where $\mathcal{P}_{\mathcal{R}}^{\mathrm{QT}}(k)$ is the quantum-theoretical
or equilibrium power spectrum. Thus measurements of $C_{l}$ may be used to set
experimental limits on $\xi(k)$ \cite{AV10}.

The function $\xi(k)$ quantifies the degree of nonequilibrium as a function of
$k$. In a model with a pre-inflationary era, extensive numerical simulations
show that $\xi(k)$ is expected to take the form of an inverse-tangent -- with
$\xi<1$ for small $k$ and $\xi\simeq1$ at large $k$ \cite{CV14}. The extent to
which this prediction is supported by the data is currently under study
\cite{PVV14}.

Incomplete relaxation in the past is one means by which nonequilibrium could
exist in the inflationary era. Another possibility is that nonequilibrium is
\textit{generated} during the inflationary phase by exotic gravitational
effects at the Planck scale (ref.\ \cite{AV10}, section IVB). Trans-Planckian
modes -- that is, modes that originally had sub-Planckian physical wavelengths
-- may well contribute to the observable part of the inflationary spectrum
\cite{BM01,MB01}, in which case inflation provides an empirical window onto
physics at the Planck scale \cite{BM13}. It has been suggested that quantum
equilibrium might be gravitationally unstable \cite{AV04b,AV07}. In quantum
field theory the existence of an equilibrium state arguably requires a
background spacetime that is globally hyperbolic, in which case nonequilibrium
could be generated by the formation and evaporation of a black hole (a
proposal that is also motivated by the controversial question of information
loss) \cite{AV04b,AV07}. A heuristic picture of the formation and evaporation
of microscopic black holes then suggests that quantum nonequilibrium will be
generated at the Planck length $l_{\mathrm{P}}$. Such a process could be
modelled in terms of nonequilibrium field modes. Thus, a mode that begins with
a physical wavelength $\lambda_{\mathrm{phys}}<l_{\mathrm{P}}$ in the early
inflationary era may be assumed to be out of equilibrium upon exiting the
Planckian regime (that is, when $\lambda_{\mathrm{phys}}>l_{\mathrm{P}}$)
\cite{AV10}. If such processes exist, the inflaton field will carry quantum
nonequilibrium at \textit{short} wavelengths (below some comoving cutoff).

For our present purpose, the main conclusion to draw is that the inflaton
field may act as a carrier of primordial nonequilibrium -- whether it is relic
nonequilibrium from incomplete relaxation during a pre-inflationary era, or
nonequilibrium that was generated by Planck-scale effects during inflation
itself. This brings us to the question: in addition to leaving a macroscopic
imprint on the CMB, could primordial nonequilibrium survive all the way up to
the present and be found in microscopic relic systems today?

\subsection{Inflaton decay}
\label{IIB}

Post-inflation, the density of any relic particles (nonequilibrium or
otherwise) from a pre-inflationary era will be so diluted as to be completely
undetectable today. However, one may consider relic particles that were
created at the end of inflation by the decay of the inflaton field itself --
where in standard inflationary scenarios inflaton decay is in fact the source
of almost all the matter present in our universe.

To discuss this, note that in pilot-wave theory it is standard to describe
bosonic fields in terms of evolving field configurations (as in our treatment
of the free scalar field in Section \ref{IIA}) whereas there are different
approaches for fermionic fields. Arguably the most straightforward pilot-wave
theory of fermions utilises a Dirac sea picture of particle trajectories
determined by a pilot wave that obeys the many-body Dirac equation
\cite{BH93,C03,CS07}. Alternatively, a formal field theory based on
anticommuting Grassmann fields may be written down \cite{AV92,AV96} but its
interpretation presents problems that remain to be addressed \cite{S10}. For
our purposes we will assume the Dirac sea model for fermions.

During the inflationary era the inflaton field $\varphi$ is approximately
homogeneous and may be written as%
\begin{align}
\varphi(\mathbf{x},t)=\phi_{0}(t)+\phi(\mathbf{x},t)\ ,
\end{align}
where $\phi_{0}(t)$ is a homogeneous part and $\phi(\mathbf{x},t)$ (or
$\phi_{\mathbf{k}}(t)$) is a small perturbation. As we have noted, during the
inflationary expansion perturbations $\phi_{\mathbf{k}}$ do not relax to
quantum equilibrium and in fact the exponential expansion of space transfers
any nonequilibrium that may exist from microscopic to macroscopic
lengthscales. The inflaton field is then a natural candidate for a carrier of
primordial quantum nonequilibrium (whatever its ultimate origin).

The process of `preheating' is driven by the homogeneous and essentially
classical part $\phi_{0}(t)$ (that is, by the $k=0$ mode of the inflaton
field) \cite{BTW06}. The inflaton is treated as a classical external field,
acting on other (quantum) fields which become excited by parametric resonance.
Because of the classicality of the relevant part of the inflaton field, this
process is unlikely to result in a transference of nonequilibrium from the
inflaton to the created particles. During `reheating', however, perturbative
decay of the inflaton can occur, and we expect that nonequilibrium in the
inflaton field will be at least to some extent transferred to its decay products.

Note that we follow the standard procedure of treating the large homogeneous
part $\phi_{0}(t)$ as a classical field and the small perturbation
$\phi(\mathbf{x},t)$ as a quantised field. This deserves some comment. In the
context of preheating, it has been argued that $\phi_{0}(t)$ arises from a
coherent state with a space-independent quantum expectation value \cite{TB90}. It
is also common to argue that the large amplitude and large occupation number
of the `zero mode' at the end of inflation justifies it being treated as a
classical field (see for example refs. \cite{BTW06} and \cite{ABCM10}).
Here we assume the standard formalism, albeit
rewritten in de Broglie-Bohm form. By construction, then, there is no
probability distribution for $\phi_{0}(t)$ (which has a classical `known'
value at all times). Whereas $\phi(\mathbf{x},t)$ has a probability
distribution, which in the standard theory is given by the Born rule and which
in de Broglie-Bohm theory can be more general. The probability distribution
for $\phi(\mathbf{x},t)$ is used to calculate the power spectrum emerging from
the inflationary vacuum. By allowing this distribution to be out of
equilibrium, new physical effects can occur in the CMB \cite{AV10}.
In contrast, because $\phi_{0}(t)$ is
treated as a classical background field with no probability distribution there
is no question of ascribing equilibrium or nonequilibrium to this part of the
field (at least not at the level of the effective description which we adopt here).\footnote{
Note that, in the standard formalism being assumed here, even at very long
wavelengths there remains a formal distinction between the large classical
homogeneous field $\phi_{0}(t)$ and modes of the small quantised field
$\phi(\mathbf{x},t)$.}

The perturbative decay of the inflaton occurs through local interactions. For
example, reheating can occur if the inflaton field $\varphi$ is coupled to a
bosonic field $\Phi$ and a fermionic field $\psi$ via an interaction
Hamiltonian density of the form%
\begin{align}
\mathcal{H}_{\mathrm{int}}=a\varphi\Phi^{2}+b\varphi\bar{\psi}\psi\ ,
\label{intn}%
\end{align}
where $a$, $b$ are constants (ref.\ \cite{PU09}, pp. 507--510). In actual
calculations, it is usual to consider only the dominant homogeneous part
$\phi_{0}$ of the field $\varphi=\phi_{0}+\phi$, and to ignore contributions
from the small perturbation $\phi$. Because the dominant homogeneous part
$\phi_{0}$ is treated essentially classically, inflaton decay bears some
resemblance to the process of pair creation by a strong classical electric
field. 

The decay particles will have physical wavelengths no greater than the
instantaneous Hubble radius, $\lambda_{\mathrm{phys}}\lesssim H^{-1}$, since
local processes cannot significantly excite super-Hubble modes (for which the
particle concept is in any case ill-defined). This standard argument -- that
super-Hubble modes are shielded from the
effects of local interactions -- is still valid in the de Broglie-Bohm
formulation since we are speaking of the time evolution of the wave functional
$\Psi$ (and of its mode decomposition) which still satisfies the usual
Schr\"{o}dinger equation. We have a nonlocal dynamical equation \eqref{the_first} for the
evolving configuration $q(t)$, but the Schr\"{o}dinger equation for $\Psi$
takes the usual form and therefore has the usual properties. Local Hamiltonian
terms in the Schr\"{o}dinger equation will be unable to excite super-Hubble
modes just as in standard quantum field theory.

How could quantum nonequilibrium exist in the decay products? There seem to be
two possible mechanisms.

First, note that the inflaton perturbation $\phi$ will also appear in the
interaction Hamiltonian (\ref{intn}). The dominant processes of particle
creation by the homogeneous part $\phi_{0}$ will necessarily be subject to
corrections from the perturbation $\phi$. If the perturbation is out of
equilibrium, the induced corrections will carry signatures of nonequilibrium
-- as will be illustrated by a simple model of field couplings in Section \ref{particle_decay}.

Second, as in any de Broglie-Bohm account of a quantum process, the final
probability distribution for the created particles will originate from the
initial probability distribution for the \textit{complete} hidden-variable
state.\footnote{In pilot-wave theory the outcome of a single quantum
measurement is determined by the complete initial configuration (together with
the initial wave function and total Hamiltonian). Over an ensemble, the
distribution of outcomes is then determined by the distribution of initial
configurations.} In this case the initial hidden-variable state will include
vacuum bosonic field configurations together with vacuum fermionic particle
configurations for the created species (assuming a Dirac-sea account of
fermions). Thus, if the relevant vacuum variables for the created species are
out of equilibrium at the beginning of inflaton decay, the created particles
will in general violate the Born rule. As we have discussed, inflaton
perturbations do not relax to equilibrium during the inflationary phase. One
may expect that the other degrees of freedom in the vacuum will show a
comparable behaviour -- in which case they could indeed be out of equilibrium
at the onset of inflaton decay, resulting in nonequilibrium for the decay products.

At least in principle, then, the particles created by inflaton decay could
show deviations from quantum equilibrium. However, subsequent relaxation can
be avoided only if the particles are created at a time after their
corresponding decoupling time $t_{\mathrm{dec}}$ (when the mean free time
$t_{\operatorname{col}}$ between collisions is larger than the expansion
timescale $t_{\exp}\equiv a/\dot{a}$) or equivalently at a temperature below
their decoupling temperature $T_{\mathrm{dec}}$. Otherwise the interactions
with other particles are likely to cause rapid relaxation.

A natural candidate to consider is the gravitino $\tilde{G}$, which arises in
supersymmetric theories of high-energy physics. In some models, gravitinos are
copiously produced by inflaton decay \cite{EHT06,KTY06,ETY07} and could make
up a significant component of dark matter \cite{T08}. (For recent reviews of
gravitinos as dark matter candidates see for example refs. \cite{EO13,C14}.)
Gravitinos are very weakly interacting and therefore in practice could not be
detected directly, but in many models they are unstable and decay into
particles that are more readily detectable. Again, as we shall see, in general
we expect any decay process to transfer quantum nonequilibrium from the
initial decaying field to the decay products. Thus, at least in principle, one
could search for deviations from the Born rule in (say) photons that are
generated by gravitino decay. However, the decay would have to take place
after the time $(t_{\mathrm{dec}})_{\gamma}$ of photon decoupling -- so that
the decay photons may in turn avoid relaxation.

It may then seem unlikely that primordial nonequilibrium could ever survive
until the present, since several stages may be required. But simple estimates
suggest that at least in principle the required constraints could be satisfied
for some models.

The unstable gravitino $\tilde{G}$ has been estimated to decouple at a
temperature $(T_{\mathrm{dec}})_{\tilde{G}}$ given by \cite{FY02}%
\begin{align}
k_{\mathrm{B}}(T_{\mathrm{dec}})_{\tilde{G}}&\equiv x_{\tilde{G}}%
(k_{\mathrm{B}}T_{\mathrm{P}})\\&\approx(1\ \mathrm{TeV})\left(  \frac{g_{\ast}%
}{230}\right)  ^{1/2}\left(  \frac{m_{\tilde{G}}}{10\ \mathrm{keV}}\right)
^{2}\left(  \frac{1\ \mathrm{TeV}}{m_{gl}}\right)  ^{2}\ ,
\end{align}
where $T_{\mathrm{P}}$ is the Planck temperature, $g_{\ast}$ is the number of
spin degrees of freedom (for the effectively massless particles) at the
temperature $(T_{\mathrm{dec}})_{\tilde{G}}$, $m_{gl}$ is the gluino mass, and
$m_{\tilde{G}}$ is the gravitino mass. For the purpose of illustration, if we
take $\left(  g_{\ast}/230\right)  ^{1/2}\sim1$ and $\left(  1\ \mathrm{TeV/}%
m_{gl}\right)  ^{2}\sim1$, then%
\begin{align}
x_{\tilde{G}}\approx\left(  \frac{m_{\tilde{G}}}{10^{3}\ \mathrm{GeV}}\right)
^{2}\ . \label{xG}%
\end{align}
If for example we take $m_{\tilde{G}}\approx100\ \mathrm{GeV}$, then
$x_{\tilde{G}}\approx10^{-2}$. Gravitinos produced by inflaton decay at
temperatures below $(T_{\mathrm{dec}})_{\tilde{G}}\equiv x_{\tilde{G}%
}T_{\mathrm{P}}$ could potentially avoid quantum relaxation. Any
nonequilibrium which they carry could then be transferred to their decay
products. If the gravitino is not the lightest supersymmetric particle, then
it will indeed be unstable. For large $m_{\tilde{G}}$ the total decay rate is
estimated to be \cite{NY06}%
\begin{align}
\Gamma_{\tilde{G}}=(193/48)(m_{\tilde{G}}^{3}/M_{\mathrm{P}}^{2})\ ,
\end{align}
where $M_{\mathrm{P}}\simeq1.2\times10^{19}\ \mathrm{GeV}$ is the Planck mass.
The time $(t_{\mathrm{decay}})_{\tilde{G}}$ at which the gravitino decays is
of order the lifetime $1/\Gamma_{\tilde{G}}$. Using the standard
temperature-time relation%
\begin{align}
t\sim(1\ \mathrm{s})\left(  \frac{1\ \mathrm{MeV}}{k_{\mathrm{B}}T}\right)
^{2}\ , \label{tT}%
\end{align}
the corresponding temperature is then%
\begin{align}
k_{\mathrm{B}}(T_{\mathrm{decay}})_{\tilde{G}}\sim(m_{\tilde{G}}%
/1\ \mathrm{GeV})^{3/2}\ \mathrm{eV}\ .
\end{align}
For example, again for the case $m_{\tilde{G}}\approx100\ \mathrm{GeV}$, the
relic gravitinos will decay when $k_{\mathrm{B}}(T_{\mathrm{decay}}%
)_{\tilde{G}}\sim1\ \mathrm{keV}$. This is prior to photon decoupling, so that
any (potentially nonequilibrium) photons produced by the decaying gravitinos
would interact strongly with matter and quickly relax to quantum equilibrium.
To obtain gravitino decay after photon decoupling, we would need
$k_{\mathrm{B}}(T_{\mathrm{decay}})_{\tilde{G}}\lesssim k_{\mathrm{B}%
}(T_{\mathrm{dec}})_{\gamma}\sim0.3\ \mathrm{eV}$, or $m_{\tilde{G}}%
\lesssim0.5\ \mathrm{GeV}$. For such small gravitino masses, decoupling occurs
at (roughly)%
\begin{align}
(T_{\mathrm{dec}})_{\tilde{G}}=x_{\tilde{G}}T_{\mathrm{P}}\approx\left(
m_{\tilde{G}}/10^{3}\ \mathrm{GeV}\right)  ^{2}T_{\mathrm{P}}\lesssim
10^{-7}T_{\mathrm{P}}\ .
\end{align}
In such a scenario, to have a hope of finding relic nonequilibrium in photons
from gravitino decay, we would need to restrict ourselves to those gravitinos
that were produced by inflaton decay at temperatures $\lesssim10^{-7}%
T_{\mathrm{P}}$.

Our considerations here are intended to be illustrative only. It may prove
more favourable to consider other gravitino decay products -- or to apply
similar reasoning to other relics from the Planck era besides the gravitino\footnote{Colin \cite{SC13} has developed the pilot-wave theory of (first-quantised) Majorana fermions and suggests that quantum nonequilibrium might survive at sub-Compton lengthscales for these systems.}.
And of course one could also consider photons that are generated by the
annihilation of relic particles as well as by their decay.

While definite conclusions must await the development of detailed and specific
models, in principle the required constraints do not seem insuperable. There
is, however, a further question we have yet to address: whether or not
relaxation will still occur even for decay particles that are decoupled.
Decoupling is necessary but not sufficient to avoid relaxation.
We may discuss this for decay particles whose physical wavelengths are sufficiently
sub-Hubble ($\lambda_{\mathrm{phys}}<<H^{-1}$) that the Minkowski limit
applies, since extensive numerical studies of relaxation have already been
carried out in this limit.
If the decay
particles are free but in quantum states that are superpositions of even
modest numbers of energy eigenstates, then rapid relaxation will occur (on
timescales comparable to those over which the wave function itself evolves)
\cite{AV92,AV01,VW05,EC06,TRV12,SC12,ACV14}. On the other hand, if the number
of energy states in the superposition is small then it is likely that
relaxation will not take place completely. It was shown in ref.\ \cite{ACV14}
that, if the relative phases in the initial superposition are chosen randomly,
then for small numbers of energy states it is likely that the trajectories
will not fully explore the configuration space, resulting in a small but
significant non-zero `residue' in the coarse-grained $H$-function --
corresponding to a small deviation from quantum equilibrium -- even in the
long-time limit. It appears that such behaviour can occur for larger numbers
of energy states as well, but will be increasingly rare the more energy states
are present in the superposition (see ref.\ \cite{ACV14} for a detailed
discussion). Decay particles will be generated with a range of effective
quantum states. For that fraction of particles whose wave functions have a
small number of superposed energy states, there is likely to be a small
residual nonequilibrium even in the long-time limit. Therefore, again, at
least in principle there seems to be no insuperable obstacle to primordial
nonequilibrium surviving to some (perhaps small) degree until the present day.

\subsection{Relic conformal vacua}
\label{IIC}

While inflaton decay will certainly create nonequilibrium particles from an
initially nonequilibrium vacuum, we have seen that there are practical
obstacles to such nonequilibrium surviving until the present day. The
obstacles do not seem insurmountable in principle, but whether a scenario of
the kind we have sketched will be realised in practice is at present unknown.
There is, however, an alternative and rather simpler scenario which appears to
be free of such obstacles. This involves considering relic nonequilibrium
field modes for the vacuum only. This has the advantage that vacuum wave
functions are so simple that no further relaxation can be generated -- any
relic nonequilibrium from earlier times will be frozen and preserved.

But how could primordial field modes remain unexcited in the post-inflationary
era? For a given field there are three mechanisms that can cause excitation:
(i) inflaton decay, (ii) interactions with other fields, and (iii) spatial
expansion. It is, however, possible to avoid each of these. Firstly, while a
field mode is in the super-Hubble regime it will in effect be shielded from
the effects of local physics and will not be subject to excitation from
perturbative interactions (with the inflaton or with other fields).\footnote{
Again, this standard argument is still valid in the de Broglie-Bohm
formulation since we are referring to the time evolution of the wave
functional only.}
 Secondly,
if during the post-inflationary radiation-dominated phase the field mode
enters the Hubble radius at a time $t_{\mathrm{enter}}$ that is later than the
decoupling time $t_{\mathrm{dec}}$ for the corresponding particle species, the
mode will remain free of interactions and continue to be unexcited (see figure
1). Finally, the effects of spatial expansion may be avoided altogether by
restricting our attention to fields that are conformally-coupled to the
spacetime metric. For example, for a massless scalar field $\phi$ with
Lagrangian density%
\begin{align}
\mathcal{L}=\frac{1}{2}\sqrt{-\,g}\left(  \,g_{\mu\nu}\partial^{\mu}%
\phi\partial^{\nu}\phi-\frac{1}{6}\,R\phi^{2}\right)
\end{align}
(where $R$ is the curvature scalar), the dynamics is invariant under a
conformal transformation $\,g_{\mu\nu}(x)\rightarrow\,\tilde{g}_{\mu\nu
}(x)=\Omega^{2}(x)\,g_{\mu\nu}(x)$, $\phi(x)\rightarrow\tilde{\phi}%
(x)=\Omega^{-1}(x)\phi(x)$, where $\Omega(x)$ is an arbitrary spacetime
function \cite{BD82,PT09}. Because a Friedmann--Lema\^{\i}tre spacetime is
conformally related to a section of Minkowski spacetime, the spatial expansion
will not create particles for a (free) conformally-coupled field. The natural
or conformal vacuum state is stable, just as in Minkowski spacetime
\cite{BD82,PT09}. Conformal invariance is however possible only for massless
fields, whether bosonic or fermionic. As examples of conformally-coupled
particle species, we may consider photons and (if they exist) massless
neutrinos and massless gravitinos.%

\begin{figure}
\input{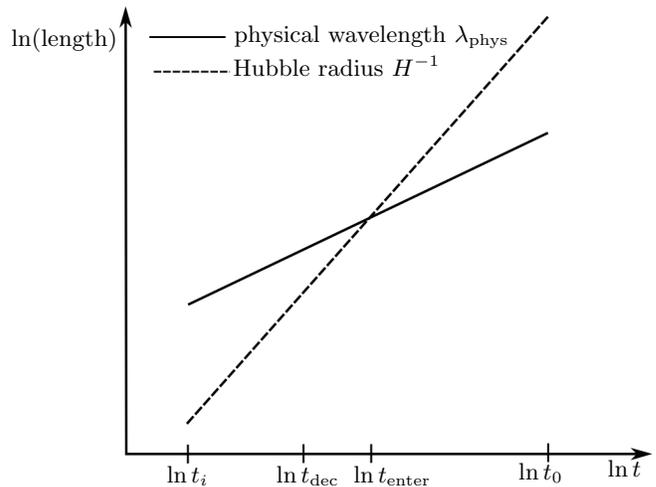}
\caption{Lengthscales for a radiation-dominated expansion. The solid line
shows the time evolution of the physical wavelength $\lambda_{\mathrm{phys}%
}=a\lambda\propto t^{1/2}$. The dashed line shows the time evolution of the
Hubble radius $H^{-1}=2t$. The mode enters the Hubble radius after the
decoupling time $t_{\mathrm{dec}}$.}%
\end{figure}

Because ground-state wave functions and the associated de Broglie velocity
fields are so simple (indeed trivial), relic vacuum modes will not relax to
equilibrium and could therefore survive as carriers of nonequilibrium until
the present day. As we shall see, nonequilibrium vacuum modes would in
principle generate corrections to particle-physics processes.

At what lengthscale might relic nonequilibrium exist in the vacuum today? This
may be estimated by requiring that the modes enter the Hubble radius at times
$t_{\mathrm{enter}}>t_{\mathrm{dec}}$ (so as to avoid excitation and hence
likely relaxation). Thus we require that at the time $t_{\mathrm{dec}}$ the
vacuum modes have an instantaneous physical wavelength $\lambda_{\mathrm{phys}%
}^{\mathrm{vac}}(t_{\mathrm{dec}})$ that is super-Hubble,%
\begin{align}
\lambda_{\mathrm{phys}}^{\mathrm{vac}}(t_{\mathrm{dec}})\gtrsim
H_{\mathrm{dec}}^{-1}\ , \label{lam-con4}%
\end{align}
where $H_{\mathrm{dec}}^{-1}$ is the Hubble radius at time $t_{\mathrm{dec}}$
(as shown in figure 1). Now $\lambda_{\mathrm{phys}}^{\mathrm{vac}%
}(t_{\mathrm{dec}})=a_{\mathrm{dec}}\lambda^{\mathrm{vac}}$ (where
$a_{\mathrm{dec}}=T_{0}/T_{\mathrm{dec}}$ and $T_{0}\simeq2.7\ \mathrm{K}$)
and $H_{\mathrm{dec}}^{-1}=2t_{\mathrm{dec}}$ with $t_{\mathrm{dec}}$
expressed in terms of $T_{\mathrm{dec}}$ by the approximate formula
(\ref{tT}). The lower bound (\ref{lam-con4}) then becomes (inserting $c$)%
\begin{align}
\lambda^{\mathrm{vac}}\gtrsim2c(1\ \mathrm{s})\left(  \frac{1\ \mathrm{MeV}%
}{k_{\mathrm{B}}T_{\mathrm{dec}}}\right)  \left(  \frac{1\ \mathrm{MeV}%
}{k_{\mathrm{B}}T_{0}}\right)
\end{align}
or%
\begin{align}
\lambda^{\mathrm{vac}}\gtrsim(3\times10^{20}\ \mathrm{cm})\left(
\frac{1\ \mathrm{MeV}}{k_{\mathrm{B}}T_{\mathrm{dec}}}\right)  \,. \label{lb2}%
\end{align}
This is a lower bound on the comoving wavelength $\lambda^{\mathrm{vac}}$ at
which nonequilibrium could be found for conformally-coupled vacuum modes.

The lower bound (\ref{lb2}) becomes prohibitively large unless we focus on
fields that decouple around the Planck temperature or soon after. For photons
$k_{\mathrm{B}}(T_{\mathrm{dec}})_{\gamma}\sim0.3\ \mathrm{eV}$, and so for
the electromagnetic vacuum (\ref{lb2}) implies $\lambda_{\gamma}%
^{\mathrm{vac}}\gtrsim10^{27}\ \mathrm{cm}$. For massless, conformally-coupled
neutrinos (if such exist), $k_{\mathrm{B}}(T_{\mathrm{dec}})_{\nu}%
\sim1\ \mathrm{MeV}$ and $\lambda_{\nu}^{\mathrm{vac}}\gtrsim10^{20}%
\ \mathrm{cm}\simeq30\ \mathrm{pc}$ (or $\sim10^{2}$ light years). Relic
nonequilibrium for these vacua could plausibly exist today only at such huge
wavelengths and any induced effects would be far beyond any range of detection
in the foreseeable future.

We must therefore consider fields that decoupled close to the Planck
temperature. Gravitons are expected to be minimally-coupled and so would not
have a stable vacuum state under the spatial expansion. However, a massless
gravitino field should be conformally-coupled, in which case it would be a
candidate for our scenario. For massless gravitinos we have a lower bound%
\begin{align}
\lambda_{\tilde{G}}^{\mathrm{vac}}\gtrsim(10^{-2}\ \mathrm{cm})(1/x_{\tilde
{G}}) \label{lbGvac}%
\end{align}
(again writing $k_{\mathrm{B}}(T_{\mathrm{dec}})_{\tilde{G}}\equiv
x_{\tilde{G}}(k_{\mathrm{B}}T_{\mathrm{P}})\simeq x_{\tilde{G}}(10^{19}%
\ \mathrm{GeV})$ and with $x_{\tilde{G}}\lesssim1$). If, for example, we take
$x_{\tilde{G}}\approx10^{-2}$ then $\lambda_{\tilde{G}}^{\mathrm{vac}}%
\gtrsim1\ \mathrm{cm}$. According to this crude and illustrative estimate,
relic nonequilibrium for a massless gravitino vacuum today appears to be
possible for modes of wavelength $\gtrsim1\ \mathrm{cm}$.

\subsection{Particle physics in a nonequilibrium vacuum}
\label{IID}

If nonequilibrium vacuum modes do exist today, how could they manifest
experimentally? In principle they would induce nonequilibrium corrections to
particle creation from the vacuum (as already noted for inflaton decay) or to
other perturbative processes such as particle decay.

Consider for example a free scalar field $\Phi(\mathbf{x},t)$ that is massive
and charged. Let us again write the Fourier components as $\Phi_{\mathbf{k}%
}(t)=\left(  \sqrt{V}/(2\pi)^{3/2}\right)  \left(  Q_{\mathbf{k}%
1}(t)+iQ_{\mathbf{k}2}(t)\right)  $ with real $Q_{\mathbf{k}r}$ ($r=1,2$). In
Minkowski spacetime -- which is suitable for a description of local laboratory
physics -- the vacuum wave functional takes the form%
\begin{align}
\Psi_{0}[Q_{\mathbf{k}r},t]\propto%
{\displaystyle\prod\limits_{\mathbf{k}r}}
\exp\left(  {-}\frac{1}{2}{{\omega}}Q_{\mathbf{k}r}^{2}\right)  \exp\left(
{-i}\frac{1}{2}{{\omega t}}\right)  \label{PsiVac}%
\end{align}
where $\omega=(m^{2}+k^{2})^{1/2}$ and $m$ is the mass associated with the
field. (On expanding space the vacuum wave functional will reduce to this form
in the short-wavelength limit.)

Let us assume that the quantum state of the field is indeed the vacuum state
(\ref{PsiVac}). Assuming for simplicity that the (putative) long-wavelength
nonequilibrium modes are uncorrelated, we may then consider a hypothetical
nonequilibrium vacuum with a distribution of the form%
\begin{align}
P_{0}[Q_{\mathbf{k}r}]\propto%
{\displaystyle\prod\limits_{\substack{\mathbf{k}r\\(k>k_{\mathrm{c}})}}}
\exp\left(  {-{\omega}}Q_{\mathbf{k}r}^{2}\right)  .%
{\displaystyle\prod\limits_{\substack{\mathbf{k}r\\(k<k_{\mathrm{c}})}}}
\rho_{\mathbf{k}r}(Q_{\mathbf{k}r})\ , \label{VacNoneq}%
\end{align}
where $\rho_{\mathbf{k}r}(Q_{\mathbf{k}r})$ is a general nonequilibrium
distribution for the mode $\mathbf{k}r$ and the wavelength cutoff
$2\pi/k_{\mathrm{c}}$ is at least as large as the relevant lower bound
(\ref{lb2}) on $\lambda^{\mathrm{vac}}$. The short wavelength modes
($k>k_{\mathrm{c}}$) are in equilibrium while the long wavelength modes
($k<k_{\mathrm{c}}$) are out of equilibrium. (The vacuum distribution $P_{0}$
is time independent because the de Broglie velocity field generated by
(\ref{PsiVac}) vanishes, $\dot{Q}_{\mathbf{k}r}=0$, since the phase of the
wave functional depends on $t$ only.)

If the field $\Phi$ is now coupled to an external and classical
electromagnetic field $\mathbf{A}_{\mathrm{ext}}$, corresponding to a
replacement $\mathbf{\nabla}\Phi\rightarrow\mathbf{\nabla}\Phi+ie\mathbf{A}%
_{\mathrm{ext}}\Phi$ in the Hamiltonian, pairs of oppositely-charged bosons
will be created from the vacuum.\footnote{The de Broglie-Bohm theory of a
charged scalar field interacting with the electromagnetic field is discussed
in refs. \cite{AV92,AVbook}.} As in our discussion of inflaton decay, the
probability distribution for the created particles originates from the initial
probability distribution $P_{0}[Q_{\mathbf{k}r}]$ for the vacuum field $\Phi$.
(There are no other degrees of freedom varying over the ensemble, since the
given classical field $\mathbf{A}_{\mathrm{ext}}$ is the same across the
ensemble.) Clearly, if $P_{0}\neq\left\vert \Psi_{0}\right\vert ^{2}$ for
long-wavelength modes, the final probability distribution for the created
particles will necessarily carry traces of the initial nonequilibrium that was
present in the vacuum. We could for example consider an interaction
Hamiltonian $e^{2}A_{\mathrm{ext}}^{2}\Phi^{\ast}\Phi$ and calculate the final
particle distribution arising from a given initial nonequilibrium vacuum
distribution of the form (\ref{VacNoneq}).

Similarly, processes of particle decay will be affected by the nonequilibrium
vacuum. Consider, for example, the decay of a particle associated with a
(bosonic or fermionic) field $\psi$ that is coupled to $\Phi$ and to a third
field $\chi$. (For bosonic fields, the decay might be induced by an
interaction Hamiltonian of the form $a\chi\Phi^{2}\psi$ where $a$ is a
coupling constant.) An initial state $\left\vert \mathbf{p}\right\rangle
_{\psi}\otimes\left\vert 0\right\rangle _{\Phi}\otimes\left\vert
0\right\rangle _{\chi}$ -- where $\left\vert \mathbf{p}\right\rangle _{\psi}$
is a single-particle state of momentum $\mathbf{p}$ for the field $\psi$ and
$\left\vert 0\right\rangle _{\Phi}$, $\left\vert 0\right\rangle _{\chi}$ are
respective vacua for the fields $\Phi$ and $\chi$ -- may have a non-zero
amplitude to make a transition\footnote{
In a de Broglie-Bohm account, the apparent `collapse' of the quantum state as
indicated by equation \eqref{transition} is only an effective description. During a standard
quantum process -- such as a measurement, a scattering experiment, or general
transition between eigenstates -- an initial packet $\psi(q,0)$ on
configuration space evolves into a superposition $\psi(q,t)=\sum_{n}c_{n}%
\psi_{n}(q,t)$ of non-overlapping packets $\psi_{n}(q,t)$. The final
configuration $q(t)$ can occupy only one `branch' -- say $\psi_{i}(q,t)$,
corresponding to the $i$th `outcome'. The motion of $q(t)$ will subsequently
be affected by $\psi_{i}(q,t)$ alone, resulting in an effective `collapse' of
the wave function. The `empty' branches still exist but no longer affect the
trajectory $q(t)$. (See, for example, chapter 8 of ref. \cite{Holl93}.)}
\begin{align}\label{transition}
\left\vert \mathbf{p}\right\rangle _{\psi}\otimes\left\vert 0\right\rangle
_{\Phi}\otimes\left\vert 0\right\rangle _{\chi}\rightarrow\left\vert
0\right\rangle _{\psi}\otimes\left\vert \mathbf{k}_{1}\mathbf{k}%
_{2}\right\rangle _{\Phi}\otimes\left\vert \mathbf{p}^{\prime}\right\rangle
_{\chi}%
\end{align}
to a final state containing two excitations of the field $\Phi$ and one
excitation of $\chi$. The final probability distribution for the outgoing
particles will originate from the initial probability distribution for all the
relevant (hidden-variable) degrees of freedom -- which in this case consist of
the relevant vacuum variables for $\Phi$ and $\chi$ together with the
variables for the field $\psi$. (Again, if $\psi$ is fermionic the associated
hidden variables may consist of particle positions in the Dirac sea
\cite{BH93,C03,CS07}.) Because all these variables are coupled by the
interaction, an initial nonequilibrium distribution (\ref{VacNoneq}) for a
subset of them (that is, for the $Q_{\mathbf{k}r}$) will generally induce
corrections to the Born rule in the final joint distribution for the
collective variables and hence for the outgoing particles. Thus, for example,
for gravitinos decaying in a nonequilibrium vacuum we would expect the decay
photons to carry traces of nonequilibrium in the probability distributions for
their outgoing momenta and polarisations.

\section{Perturbative transfer of nonequilibrium}
\label{particle_decay}
We now turn to some simple but illustrative field-theoretical models of the behaviour of nonequilibrium systems. The first question that needs to be addressed is the perturbative transfer of nonequilibrium from one field to another. In this section we present a simple (bosonic) field-theoretical model that illustrates this process.

Suppose we have two Klein-Gordon fields $\phi_1$ and $\phi_2$, confined inside a box of volume $V$ with dimensions $l_x$, $l_y$, and $l_z$ such that the fields are necessarily zero valued on the boundaries of the box. In consideration of these boundary conditions, we expand and quantise the fields in a set of standing waves as $(i=1,2)$
\begin{align}
\phi_i(\mathbf{x})&=\sum_{\mathbf{k}}\frac{2^{3/2}q_{i\mathbf{k}}}{\sqrt{V}}\sin(k_x x)\sin(k_y y)\sin(k_z z),
\end{align}
with annihilation operators
\begin{align}
a_{i\mathbf{k}}&=\sqrt{\frac{\omega_{i\mathbf{k}}}{2}}\left(q_{i\mathbf{k}}+\frac{i}{\omega_{i\mathbf{k}}}p_{i\mathbf{k}}\right),
\end{align}
and a total Hamiltonian
\begin{align}
H_0&=\sum_{\mathbf{k}}\left(\omega_{1\mathbf{k}}a^\dagger_{1\mathbf{k}}a_{1\mathbf{k}}+\omega_{2\mathbf{k}}a^\dagger_{2\mathbf{k}}a_{2\mathbf{k}}\right).
\end{align}
We have dropped the zero point energy, and $k_x=n\pi/l_x$ and similarly for $y$ and $z$.  The two fields are coupled by the interaction Hamiltonian
\begin{align}
H_{\text{I}}&=g\int_V\mathrm{d}^3x\phi_1(\mathbf{x})\phi_2(\mathbf{x})\label{quad_int_ham}\\
&=\frac{g}{2}\sum_\mathbf{k}\frac{1}{\sqrt{\omega_{1\mathbf{k}}\omega_{2\mathbf{k}}}}(a_{1\mathbf{k}}+a^\dagger_{1\mathbf{k}})(a_{2\mathbf{k}}+a^\dagger_{2\mathbf{k}}),
\end{align}
where $g$ is a coupling constant.
If we suppose that at time $t=0$ the system is in the free (unperturbed) eigenstate $\ket{E_i}$, the first order perturbative amplitude to transition to the state $\ket{E_f}$ is 
\begin{align}
d_f^{(1)}(t)=\bra{f}H_{\text{I}}\ket{i}\frac{e^{-iE_ft}-e^{-iE_it}}{E_f-E_i}.\label{amp}
\end{align}
This will be damped for any $E_f$ significantly different from $E_i$. We may exploit this fact by further insisting that
\begin{itemize}
\item $l_x\gg l_y \gg l_z$, so that the lowest mode of each field is significantly lower than all others, and
\item the limit $m_2\rightarrow m_1$ is taken, so that the lowest modes of $\phi_1$ and $\phi_2$ have the same unperturbed energy.
\end{itemize}
These conditions ensure that the system state in which field $\phi_1$ has one particle occupying its lowest mode and the field $\phi_2$ is a vacuum has identical unperturbed energy to the system state in which the individual field states are reversed. We shall denote these states $\ket{1,0}$ and $\ket{0,1}$ respectively. Since these states have identical unperturbed energies, the first order perturbative amplitudes \eqref{amp} between the states is significantly amplified, whereas all others damped. This is the justification of the rotating wave approximation, familiar from quantum optics and cavity QED (see for instance ref.\ \cite{knight}). Put simply, the states $\ket{1,0}$ and $\ket{0,1}$ are strongly coupled to each other and only very weakly coupled to any other state. 

We make the rotating wave approximation by removing all terms in the Hamiltonian that would effect an evolution to states other than $\ket{1,0}$ and $\ket{0,1}$. This allows us to employ the effective Hamiltonian,
\begin{align}
H_{\text{eff}}=\omega(a_1^\dagger a_1 + a_2^\dagger a_2)+\frac{g}{2\omega}(a_1a_2^\dagger + a_2a_1^\dagger).
\end{align}
We have suppressed the mode subscripts for simplicity. The approximate Schr\"{o}dinger equation $H_{\text{eff}}\ket{\psi}=i\partial_t\ket{\psi}$, along with the initial condition $\left.\ket{\psi}\right|_{t=0}=\ket{1,0}$, yields the solution 
\begin{align}
\ket{\psi}=e^{-i\omega t}\left(\cos\left(\frac{gt}{2\omega}\right)\ket{1,0}-i\sin\left(\frac{gt}{2\omega}\right)\ket{0,1}\right)\label{part_decay_state}.
\end{align}
The sine and cosine in Eq.\ \eqref{part_decay_state} describe an oscillatory decay process in which the first type of particle is seen to decay into the second type, which promptly decays back. This type of flip-flopping between one type of particle and the other is functionally equivalent to vacuum-field Rabi oscillations in the Jaynes-Cummings model \cite{knight,jc} of quantum optics and cavity QED wherein an exchange of energy occurs between an atom and a cavity mode of the electromagnetic field, perpetually creating a photon, then destroying it only to create it once more.\footnote{From a field-theoretical viewpoint the quadratic interaction \eqref{quad_int_ham}
may seem too trivial an example since the interaction may be removed by a
linear transformation of the field variables. Such a transformation would not,
however, negate the physical meaning of the original system. Our aim is to
illustrate with a simple example how nonequilibrium may be passed from one
type of field to another. We expect a similar passing of quantum
nonequilibrium between fields to be caused by any reasonable interaction term.
Our example is based on a model -- widely used in quantum optics to study the
interaction between a two-level atom and a single mode of the quantised
electromagnetic field inside a cavity -- that is simple enough to be tractable
while at the same time providing a genuine field-theoretical account of energy
transfer to and from a quantised field.}

\begin{figure}
\input{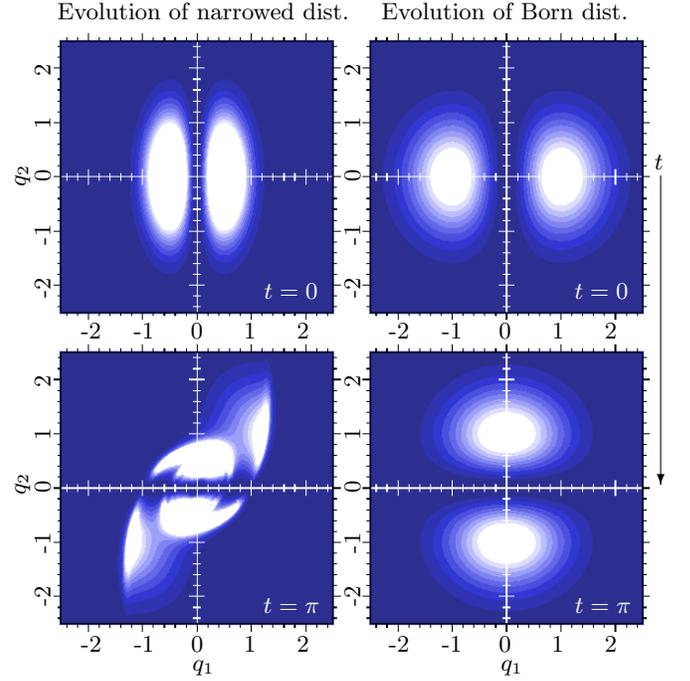}
\caption{\label{part_decay_non-eq}
The evolution of quantum equilibrium and nonequilibrium through the particle decay process described by state \eqref{part_decay_state} and guidance equations \eqref{part_guidance_part}. Initially state \eqref{part_decay_state} is a product between an excited (one particle) state in $q_1$ and a ground (vacuum) state in $q_2$. This is shown in the top right graph. As time passes, $t=0\rightarrow\pi$, the excited state in $q_1$ decays into exactly the same excited state in $q_2$. At time $t=\pi$ the state \eqref{part_decay_state} exists in another product state, except this time with excited and ground states switched between fields. This is shown in the bottom right graph. The evolution of a quantum nonequilibrium distribution is shown in the left column. Before the interaction takes place, quantum nonequilibrium exists only in the one particle state of the first field; it has been narrowed with respect to the equilibrium distribution. As time passes, the first field generates nonequilibrium in the second. At $t=\pi$, by standard quantum mechanics, the decay process is complete and there exists another product state. In contrast, the introduction of quantum nonequilibrium has created a distribution at $t=\pi$ that is correlated between $q_1$ and $q_2$. The marginal distributions for the fields are shown in figure \ref{part_marginals}. (This figure takes $\omega=g=1$.)}
\end{figure}

\begin{figure}
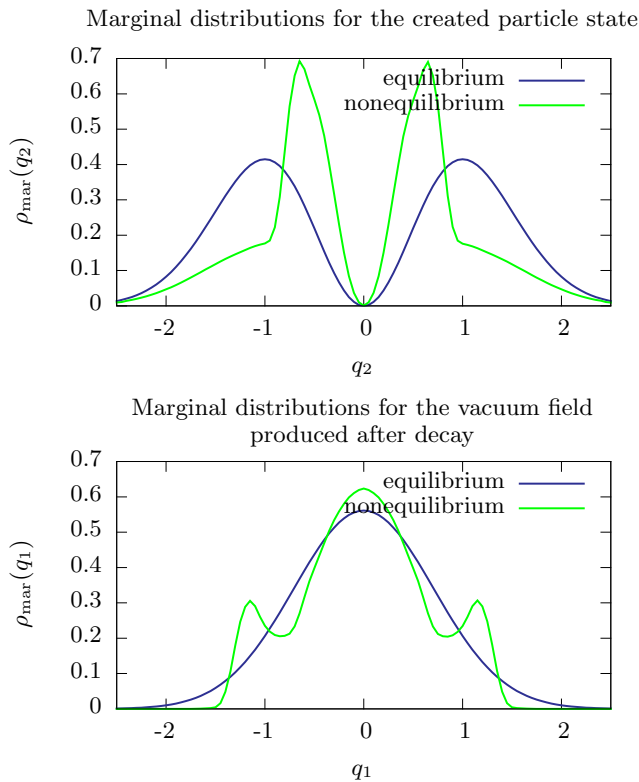

    \input{marginals_particle_created}
    \input{marginals_particle_destroyed}
\caption{\label{part_marginals}The ensemble marginal distributions of the particle and vacuum state created at $t=\pi$ in the decay process shown in figure \ref{part_decay_non-eq}. The top graph shows the marginal distribution for the excited field $\rho_{\text{mar}}(q_2)=\left.\int \mathrm{d}q_1\rho(q_1,q_2,t)\right|_{t=\pi}$. The bottom graph shows the nonequilibrium marginal distribution of the vacuum field $\rho_{\text{mar}}(q_1)=\left.\int \mathrm{d}q_2\rho(q_1,q_2,t)\right|_{t=\pi}$, obtained after the original particle state has decayed.}
\end{figure}

To develop a de Broglie-Bohm description of this particle decay process, one needs to specify the configuration of the system. For bosonic fields the canonical approach \cite{B52b} is to use the Schr\"{o}dinger representation with mode amplitudes as the configuration. In our case this is particularly simple; the state of any one system in an ensemble is described by the coordinates $q_1$ and $q_2$, proportional to the amplitudes of the lowest (standing) mode of each field. In this representation the Hamiltonian is
\begin{align}
H_{\text{eff}}=&-\frac{1}{2}\left(\partial_{q_1}^2+\partial_{q_2}^2\right)+\frac12\omega^2\left(q_1^2+q_2^2\right)\nonumber\\
&-\omega+\frac{g}{2}\left(q_1q_2-\frac{1}{\omega^2}\partial_{q_1}\partial_{q_2}\right).\label{coord_rwa_ham}
\end{align}
In the rotating wave approximation there are derivative terms in the interaction Hamiltonian. The de Broglie velocity fields associated with the Hamiltonian \eqref{coord_rwa_ham} may be derived in the standard way, and by using 
\begin{align}
&\psi^*\partial_{q_1}\partial_{q_2}\psi-\psi\partial_{q_1}\partial_{q_2}\psi^*\nonumber\\
&=i\partial_{q_1}\left(|\psi|^2\partial_{q_2}\text{Im}\ln\psi\right)+i\partial_{q_2}\left(|\psi|^2\partial_{q_1}\text{Im}\ln\psi\right)
\end{align}
(a special case of the general identity 2 of ref.\ \cite{SV08}). Writing $\psi=|\psi|e^{iS}$, the guidance equations may be expressed as
\begin{equation}\label{part_guidance}
\begin{aligned}
\dot{q_1}&=\partial_{q_1}S+\frac{g}{2\omega^2}\partial_{q_2}S,\\
\dot{q_2}&=\partial_{q_2}S+\frac{g}{2\omega^2}\partial_{q_1}S.
\end{aligned}
\end{equation}
For the particular state \eqref{part_decay_state}, these yield
\begin{equation}\label{part_guidance_part}
\begin{aligned}
\dot{q_1}&=\frac{\frac12\left(q_2-\frac{g}{2\omega^2}q_1\right)\sin\left(\frac{gt}{\omega}\right)}{q_1^2\cos^2\left(\frac{gt}{2\omega}\right)+ q_2^2\sin^2\left(\frac{gt}{2\omega}\right)},\\
\dot{q_2}&=\frac{\frac12\left(-q_1+\frac{g}{2\omega^2}q_2\right)\sin\left(\frac{gt}{\omega}\right)}{q_1^2\cos^2\left(\frac{gt}{2\omega}\right)+ q_2^2\sin^2\left(\frac{gt}{2\omega}\right)}.
\end{aligned}
\end{equation}
The configuration $q(t)=(q_1(t),q_2(t))$ and velocity $\dot{q}(t)=(\dot{q_1}(t),\dot{q_2}(t))$ of a particular member of an ensemble evolving along a trajectory described by Eqs.\ \eqref{part_guidance_part} has the properties  
$q(t)=q(t+2\pi\omega/g)$, $\dot{q}(t)=\dot{q}(t+2\pi\omega/g)$, $q(t)=q(-t)$, and $\dot{q}(t)=-\dot{q}(-t)$.
The trajectories $q(t)$ are periodic, and halfway through their period backtrack along their original path.

Given the velocity field \eqref{part_guidance_part}, we may integrate the continuity equation \eqref{ant_cont} to obtain the time evolution of an arbitrary distribution $\rho$. (Our numerical method is described in the appendix.)

In figure \ref{part_decay_non-eq} we compare the evolution of quantum nonequilibrium with that of equilibrium for the case $\omega=g=1$. The decay from an initial product state $\ket{1,0}$ to a final product state $\ket{0,1}$ is seen in the (product) equilibrium distributions on the right-hand side of figure \ref{part_decay_non-eq}. We illustrate the transfer of nonequilibrium in the left-hand side of figure \ref{part_decay_non-eq} for the case of an initial nonequilibrium that has simply been narrowed in $q_1$ (with respect to equilibrium). Hence only the first field is initially out of equilibrium. As time passes the distribution becomes correlated in $q_1$ and $q_2$. At $t=\pi$, when according to standard quantum mechanics we should find another product state (corresponding to $\ket{0,1}$), there exists a complicated overall nonequilibrium in $(q_1,q_2)$. The marginal distributions are shown in figure \ref{part_marginals}. 

The evolution of nonequilibrium depends strongly on the particular values of $\omega$ and $g$, although in general we see two important properties of this evolution. Firstly it is apparent from figures \ref{part_decay_non-eq} and \ref{part_marginals} that nonequilibrium in the marginal distribution of the original particle state (or excited field) will generate nonequilibrium in its decay product. Secondly, although the initial product state $\left.\ket{\psi}\right|_{t=0}=\ket{1,0}$ evolves into the product state $\ket{0,1}$ at $t=\pi\omega/g$, the nonequilibrium distribution is correlated between the two fields. Such correlation could not exist in standard quantum mechanics.

\section{Energy measurements and nonequilibrium spectra}
\label{IV}
In this section we focus on quantum-mechanical measurements of energy for elementary field-theoretical systems in nonequilibrium.
As we have discussed, in this paper we restrict ourselves to simple models that may be taken to illustrate some of the basic phenomena that could occur. 

The following analysis is presented for the electromagnetic field, partly because it provides a convenient  illustrative model and partly because (as explained in Section \ref{II}) we envisage the possibility of detecting decay photons produced by particles in nonequilibrium rather than the parent particles themselves. However, the analysis should apply equally well to other field theories. 
\subsection{Setup and effective wave function}\label{setup}
\begin{figure*}[t]
\input{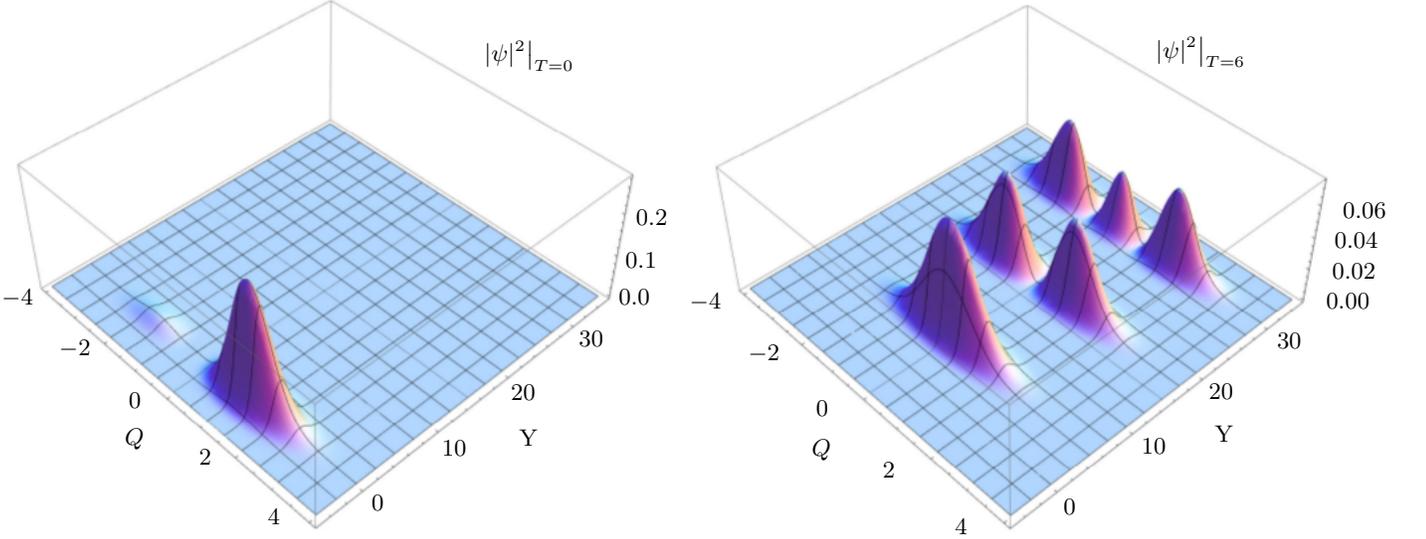}
\caption{\label{fig_1}Illustration of the energy measurement process, showing the evolution of the Born distribution into disjoint packets (for the case $c_0=c_1=c_2=1/\sqrt{3}$).
The variables $q$, $y$ and $t$ have been replaced by the rescaled variables $Q$, $Y$ and $T$ defined in section \ref{non_dim}.
The initial pointer wave function is chosen to be a Gaussian centred on Y=0.
Initially the components of the total wave function overlap and interfere. As time passes each component moves in the $Y$ direction with speed $2n+1$. After some time the components no longer overlap and the experimenter may unambiguously read off the energy eigenvalue from the position of the pointer ($Y$ coordinate).}
\end{figure*}
We work in the Coulomb gauge, $\nabla.\mathbf{A}(\mathbf{x},t)=0$, with the field expansion
\begin{align}
\mathbf{A}(\mathbf{x},t)=\sum_{\mathbf{k'}s'}\left[A_{\mathbf{k'}s'}(t)\mathbf{u}_{\mathbf{k'}s'}(\mathbf{x})+A_{\mathbf{k'}s'}^*(t)\mathbf{u}_{\mathbf{k'}s'}^*(\mathbf{x})\right],\label{u_expansion}
\end{align}
where the functions
\begin{align}
\mathbf{u}_{\mathbf{k'}s'}(\mathbf{x})=\frac{\bm{\varepsilon}_{\mathbf{k'}s'}}{\sqrt{2\varepsilon_0 V}}e^{i\mathbf{k'.x}}
\end{align}
and their complex conjugates define a basis for the function space. 
In expansion \eqref{u_expansion} and henceforth, summations over wave vectors are understood to extend over half the possible values of $\mathbf{k}'$. This is to avoid duplication of bases $\mathbf{u}^*_{\mathbf{k'}s'}$ with $\mathbf{u}_{\mathbf{-k'}s'}$. See for instance reference \cite{schiff}. The primes are included for later convenience. This expansion allows one to write the energy of the electromagnetic field as
\begin{align}
U&=\frac12 \int_V\mathrm{d}^3x\left(\varepsilon_0 \mathbf{E.E}+\frac{1}{\mu_0}\mathbf{B.B}\right)\\
&=\sum_{\mathbf{k'}s'}\frac{1}{2}\left(\dot{A}_{\mathbf{k'}s'}\dot{A}_{\mathbf{k'}s'}^*+\omega_\mathbf{k'}^2A_{\mathbf{k'}s'}A_{\mathbf{k'}s'}^*\right),\label{comp_HOs}
\end{align}
where $\omega_\mathbf{k'}=c|\mathbf{k'}|$.
Equation \ref{comp_HOs} defines a decoupled set of complex harmonic oscillators of unit mass. We shall prefer instead to work with real variables and so we decompose $A_{\mathbf{k}'s'}$ into its real and imaginary parts 
\begin{align}
A_{\mathbf{k'}s'}=q_{\mathbf{k'}s'1} +iq_{\mathbf{k'}s'2}.
\end{align}
One may then write the free field Hamiltonian as
\begin{align}
H_0=\sum_{\mathbf{k'}s'r'}H_{\mathbf{k'}s'r'}
\end{align}
with $r'=1,2$, where
\begin{align}
H_{\mathbf{k'}s'r'}=\frac{1}{2}\left(p_{\mathbf{k'}s'r'}^2+\omega_{\mathbf{k'}}^2q_{\mathbf{k'}s'r'}^2\right),
\end{align}
and where $p_{\mathbf{k'}s'r'}$ is the momentum conjugate of $q_{\mathbf{k'}s'r'}$.

Suppose we wish to perform a quantum energy measurement for a single mode of the field. We may follow the pilot-wave theory of ideal measurements described in ref.\ \cite{Holl93}. The system is coupled to an apparatus pointer with position variable $y$. The interaction Hamiltonian $H_{\text{I}}$ is taken to be of the form $g \hat{\omega}p_y$, where again $g$ is a coupling constant and $\hat{\omega}$ is the operator corresponding to the observable to be measured. In our case we have
\begin{align}\label{coupling}
H_{\text{I}}=g H_{\mathbf{k}sr}p_y,
\end{align}
where $p_y$ is the momentum conjugate to the pointer position $y$
and where $\mathbf{k}$, $s$ and $r$ refer specifically to the field mode that is being measured. Including the free Hamiltonian $H_{\text{app}}$ for the apparatus, the total Hamiltonian is
\begin{align}
H_\text{tot}=H_0+H_\text{app}+H_{\text{I}}.\label{hamiltonian}
\end{align}
We assume an initial product state 
\begin{align}
\psi(0)=\psi_{\mathbf{k}sr}(q_{\mathbf{k}sr},0)\phi(y,0)\chi(\mathcal{Q},0),
\end{align}
where $\psi_{\mathbf{k}sr}$ is the wave function for the mode in question, $\phi$ is the apparatus wave function and $\chi$ is a function of the rest of the field variables $\mathcal{Q}=\{q_{\mathbf{k'}s'r'}|(\mathbf{k'},s',r')\neq (\mathbf{k},s,r)\}$. The function $\chi$ is left unspecified as there is no need to make assumptions concerning the state of the rest of the field. Now, since $H_{\text{I}}$ and $H_\text{app}$ commute with all the terms in $H_0$ that include $\mathcal{Q}$, under time evolution the $\chi$ function remains unentangled with the rest of the system while the apparatus and the mode being measured become entangled. We may then write
\begin{align}
\psi(t)&=\Psi(q_{\mathbf{k}sr},y,t)\chi(\mathcal{Q},t),
\end{align}
where
\begin{equation}
\begin{aligned}
\Psi(q_{\mathbf{k}sr},y,t)&=\exp\left[-i(H_{\mathbf{k}sr}+H_\text{app}+gH_{\mathbf{k}sr}p_y)t\right]\\
&\times\psi_{\mathbf{k}sr}(q_{\mathbf{k}sr},0)\phi(y,0),\\
\chi(\mathcal{Q},t)&=\left[\prod_{(\mathbf{k'}s'r')\neq(\mathbf{k}sr)}\exp\left(-iH_{\mathbf{k'}s'r'}t\right)\right]\chi(\mathcal{Q},0).
\end{aligned}
\end{equation}
Since the system and apparatus remain unentangled with $\chi$, the dynamics remain completely separate. We may concern ourselves only with $\Psi(q_{\mathbf{k}sr},y,t)$ as an effective wave function. The velocity field in the ($q_{\mathbf{k}sr},y)$ plane depends on the position in that plane but is independent of the position in $\mathcal{Q}$. We may then omit the $\mathbf{k}sr$ labels in $q_{\mathbf{k}sr}$ and $H_{\mathbf{k}sr}$, and the $\mathbf{k}$ label in $\omega_\mathbf{k}$.

Let the measurement process begin at $t=0$ when the coupling is switched on. 
As usual in the description of an ideal von Neumann measurement (see for example ref.\ \cite{Holl93}), we take $g$ to be so large that the free parts of the Hamiltonian may be neglected during the measurement.
The system will then evolve according to the Schr\"{o}dinger equation
\begin{align}
\left(\partial_t+gH\partial_y\right)\Psi=0.\label{sch1}
\end{align}
Expanding $\Psi$ in a basis $\psi_n(q)$ of energy states for the field mode, we have the solution
\begin{align}
\Psi(q,y,t)=\sum_nc_n\phi(y-gE_nt)\psi_n(q).\label{key}
\end{align}
where we choose $\phi$ and $\psi_n$ to be real and $\sum_n|c_n|^2=1$.
Equation \eqref{key} describes the measurement process. If the initial system state is an energy eigenstate ($c_n=\delta_{mn}$ for some $m$), the pointer packet is translated with a speed proportional to the energy $E_m$ of the eigenstate. By observing the displacement of the pointer after a time $t$, an experimenter may infer the energy of the field mode. If instead the field mode is initially in a superposition of energy states, the different components of the superposition will be translated at different speeds until such a time when they no longer overlap and thus do not interfere. An example of this evolution into non-overlapping, non-interfering packets is shown in figure \ref{fig_1}. At this time an experimenter could unambiguously read off an energy eigenvalue. The weightings $|c_n|^2$ in the superposition could be determined by readings over an ensemble.

\subsection{Pointer packet and rescaling}\label{non_dim}
For simplicity we choose the initial pointer wave function $\phi$ in Eq.\ \eqref{key} to be a Gaussian centred on $y=0$, 
\begin{align}
\phi(y)&=\sigma^{-\frac12}(2\pi)^{-\frac14}e^{-y^2/4\sigma^2},
\end{align}
where $\sigma^2$ is the variance of $|\phi(y)|^2$.

It is convenient to introduce the rescaled parameters
\begin{align}\label{coords}
Q=\sqrt{\omega}q,\quad Y=\frac{y}{\sigma}, \quad T=\frac{g \omega t}{2\sigma}.
\end{align}
The evolution of the wave function is then determined by the Schr\"{o}dinger equation,
\begin{align}
\partial_T\Psi=\left(\partial_Q^2-Q^2\right)\partial_Y\Psi.\label{schro}
\end{align}
The general solution is
\begin{align}\label{wvfn}
&\Psi(Q,Y,T)=\sum_n\frac{c_n}{\sqrt{\pi 2^{n+1/2}n!}}\nonumber\\
&\times\exp\left[-\frac14\left(Y-(2n+1)T\right)^2\right]e^{-Q^2/2}H_n(Q),
\end{align}
where $H_n(Q)$ are Hermite polynomials.
(Equation \eqref{wvfn} differs from Eq.\ \eqref{key} by a factor $\sigma^{1/2}\omega^{-1/4}$, to normalise the wave function in the rescaled configuration space.)

\subsection{Continuity equation and guidance equations}
\label{sec:cont_eq}
From the Schr\"{o}dinger equation \eqref{schro}, it is simple to arrive at
\begin{align}
\partial_T|\Psi|^2&=\Psi^*\partial^2_Q\partial_Y\Psi+\Psi\partial^2_Q\partial_Y\Psi^*
-\partial_Y\left(Q^2|\Psi|^2\right).
\end{align}
From here we use the identity
\begin{align}
&\Psi^*\partial_Q^2\partial_Y\Psi+\Psi\partial_Q^2\partial_Y\Psi^*\nonumber\\
\equiv&\frac13\partial_Q\left(2\Psi\partial_Q\partial_Y\Psi^*-\partial_Y\Psi\partial_Q\Psi^*\right.\nonumber\\
&\left.-\partial_Q\Psi\partial_Y\Psi^*+2\Psi^*\partial_Q\partial_Y\Psi\right)\nonumber\\
&+\frac13\partial_Y\left(\Psi\partial^2_Q\Psi^*-\partial_Q\Psi\partial_Q\Psi^*+\Psi^*\partial_Q^2\Psi\right).
\end{align}
This, again, is a special case of the general identity 2 of \cite{SV08}. The continuity equation is found to be
\begin{align}
\partial_T|\Psi|^2+\partial_Q\text{Re}\left(\frac23\partial_Y\Psi\partial_Q\Psi^*-\frac43\Psi^*\partial_Q\partial_Y\Psi \right)&\nonumber\\
+\partial_Y\text{Re}\left(\frac13|\partial_Q\Psi|^2-\frac23\Psi^*\partial^2_Q\Psi+Q^2\right)&=0,
\end{align}
from which we may deduce the de Broglie guidance equations
\begin{align}
\partial_TQ&=\text{Re}\left(-\frac43\frac{\partial_Q\partial_Y\Psi}{\Psi}+\frac23\frac{\partial_Y\Psi\partial_Q\Psi^*}{|\Psi|^2}  \right),\label{guidance_a}\\
\partial_TY&=\text{Re}\left(-\frac{2}{3}\frac{\partial^2_Q\Psi}{\Psi}+\frac{1}{3}\frac{\partial_Q\Psi\partial_Q\Psi^*}{|\Psi|^2}\right)+Q^2.\label{guidance_b}
\end{align}
The factor $Q^2$ in Eq.\ \eqref{guidance_b} will turn out to have the most predictable effect on the evolution of quantum nonequilibrium in section \ref{results}. Any individual system in which $|Q|$ is abnormally large will, at least to begin with, have an abnormally large pointer velocity. The $Q^2$ term originates from the potential term in $H_{\mathbf{k}sr}=\frac{1}{2}p_{\mathbf{k}sr}^2+\frac12 \omega_{\mathbf{k}}^2q_{\mathbf{k}sr}^2$. 

\begin{figure*}[t]
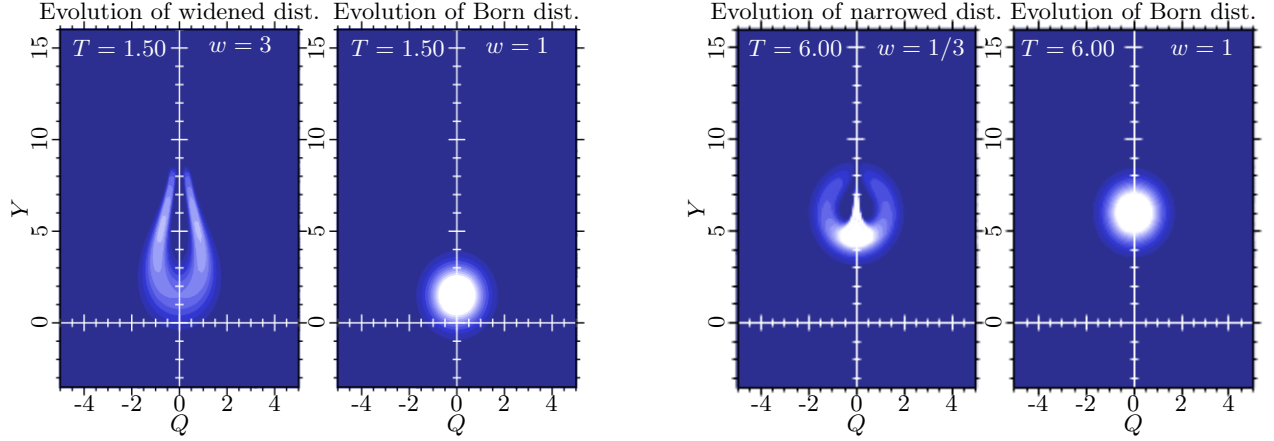

\input{gs_widened.eps_tex}
\input{gs_narrowed.eps_tex}
\caption{\label{ground_sims}The evolution of vacuum nonequilibria under an energy measurement process (as simulated by the code discussed in the appendix). On the left is a snapshot of the evolution of a widened initial $\rho$ with $w=3$, taken at $T=1.50$. The tails of $\rho$ evolve quickly to large $Y$ and small $Q$. These tails are evident in the marginal distribution for $Y$ shown in figure \ref{marginals}. On the right is the same simulation except narrowed by a factor $w=1/3$, and taken at $T=6.00$. In this case $\rho$ remains in what might loosely be deemed the support of $|\Psi|^2$, though displaying internal structure. The narrowed $\rho$ initially lags behind the Born distribution, before getting swept outwards and upwards, creating a double-bump in the pointer marginal distribution. Note that the equilibrium pointer distribution undergoes an upward displacement to indicate the zero-point energy of the vacuum mode.}
\end{figure*}

In contrast with section \ref{particle_decay}, here we have chosen to retain the zero-point energy of the $\mathbf{k}sr$ mode. Since the pointer is coupled to the total energy of the $\mathbf{k}sr$ mode, this does affect the dynamics though only in a minor respect. Had we normal ordered Eq.\ \eqref{hamiltonian}, the pointer velocity Eq.\ \eqref{guidance_b} would have an extra additive term of $-1$. In the state-specific expressions of section \ref{state_specific}, normal ordering is equivalent to switching to a coordinate system moving in the $+Y$ direction at a (rescaled) velocity of $1$, the velocity of the vacuum component in Eq.\ \eqref{wvfn}. Equivalently, one may use the coordinate transformation $Y\rightarrow Y'=Y-T$, which we shall indeed do in section \ref{long_time}.

\subsection{Expressions for three examples}\label{state_specific}
\subsubsection{Vacuum}
If the field mode being measured is in its vacuum state ($c_n=\delta_{n0}$), the evolution of the total wave function \eqref{wvfn} and the associated velocity fields \eqref{guidance_a} and \eqref{guidance_b} are given by 
\begin{align}
\Psi&=2^{-\frac14}\pi^{-\frac12}\exp\left(-\frac12 Q^2 -\frac14(Y-T)^2 \right),\\
\partial_TQ&=\frac13 Q(T-Y),\label{gs_q_vel}\\
\partial_TY&=\frac{2}{3}(1+Q^2).\label{gs_y_vel}
\end{align}
\subsubsection{One particle state}
If instead the field mode being measured contains one particle or excitation ($c_n=\delta_{n1}$), the relevant expressions are
\begin{align}
\Psi&=2^{\frac14}\pi^{-\frac12}Q\exp\left(-\frac12 Q^2-\frac14 (Y-3T)^2\right),\label{sp_measurement}\\
\partial_TQ&=\frac13 (Y-3T)\left(\frac1Q -Q\right)\label{e1},\\
\partial_TY&=\frac13\frac{1}{Q^2}+\frac43 +\frac23 Q^2\label{e2}.
\end{align}
\subsubsection{Initial superposition of vacuum and one particle state}
For a superposition, the relative phases in the $c_n$'s will contribute to the dynamics. For a superposition of initial vacuum and one particle states, we take $c_0=e^{i\theta}/\sqrt{2}$ and $c_1=1/\sqrt{2}$. Our expressions then become
\begin{widetext}
\begin{align}
\Psi&=\left(\frac{e^{i\theta}}{\sqrt{2}}+Qe^{T(Y-2T)}\right)\label{super1}
2^{-\frac14}\pi^{-\frac12}e^{-\frac14\left(Y-T\right)^2}
\exp\left(-\frac12 Q^2\right),\\
\partial_TQ&=\text{Re}\left(\frac{-\frac53T+\frac23Q^2T+\frac13Y}{\frac{e^{i\theta}}{\sqrt{2}}e^{T(2T-Y)}+Q}\right)+\frac{\frac23 QT}{\left|\frac{e^{i\theta}}{\sqrt{2}}e^{T(2T-Y)}+Q\right|^2}-\frac13(Y-T)Q\label{super2},\\
\partial_TY&=\text{Re}\left[\frac{\frac23Q}{\frac{e^{i\theta}}{\sqrt{2}}e^{T(2T-Y)}+Q}\right]+\frac{\frac13}{\left|\frac{e^{i\theta}}{\sqrt{2}}e^{T(2T-Y)}+Q\right|^2}+\frac23(Q^2+1)\label{super3}.
\end{align}
\end{widetext}

\subsection{Results for nonequilibrium energy measurements}\label{results}
We will now consider outcomes of quantum energy measurements for nonequilibrium field modes. Like many features of quantum mechanics, the usual statistical energy conservation law emerges in equilibrium. But for nonequilibrium states there is no generally useful notion of energy conservation\footnote{The fundamental dynamical equation \eqref{the_first} is first-order in time and has no naturally conserved energy. When rewritten in second order form there appears a time-dependent `quantum potential' that acts as an effective external energy source \cite{Holl93}.}.

We may consider a parameterisation of nonequilibrium that simply varies the width of the Born distribution (as discussed in Section IIA for primordial perturbations). Our initial $\rho$ is written
\begin{align}
\rho(Q,Y,0)=\frac{1}{w}\left|\Psi(Q/w,Y,0)\right|^2,
\end{align}
where $w$ is a widening parameter equal to the initial standard deviation of $\rho$ relative to $|\Psi|^2$,
\begin{align}
 w=\left.\frac{\sigma_\rho}{\sigma_{|\Psi|^2}}\right|_{t=0}.
\end{align}
(Comparing with eqn.\ \eqref{ksi}, we would have $w = \sqrt{\xi}$ for primordial perturbations.)
\subsubsection{Short-time measurement of vacuum modes}\label{short_time}
\begin{figure}
\input{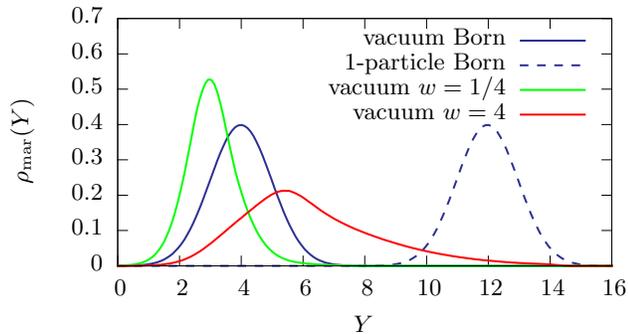}
\caption{\label{marginals}Marginal pointer distributions $\rho_{\text{mar}}(Y)$ under the energy measurement of vacuum mode nonequilibria at $T=4$. (For comparison we also show the Born pointer marginals for the vacuum and 1-particle cases.) For the widened vacuum mode ($w=4$), there is a significant probability of `detecting a particle' (that is, an excited state) in the vacuum mode. For this case there also exists a significant probability of finding the pointer around Y = 8 (which for all practical purposes would be impossible without nonequilibrium for any initial superposition). For the narrowed nonequilibrium ($w=1/4$), the pointer distribution lags behind the Born pointer distribution initially. As time progresses, $\rho$ will get swept outwards and upwards (cf.\ the right-hand side of figure \ref{ground_sims}), creating a double-bump in the pointer distribution. }
\end{figure}

In figure \ref{ground_sims} we show the short-time behaviour of widened and narrowed nonequilibrium distributions $\rho$ under the energy measurement of a vacuum mode. As the $Q^2$ term in the $Y$ velocity \eqref{gs_y_vel} dominates for any $|Q|_{t=0}>1$, widened distributions show more initial movement of the pointer. The tails of widened distributions `flick' forwards and inwards, and then seem to linger. It is at this time that the pointer position could indicate the detection of an excited state (or ‘particle’) for the vacuum, or even occupy a position disallowed by standard quantum mechanics for any initial superposition of energy states (see figure \ref{marginals}). The closer the tails get to the $Q$-axis, the slower the pointer travels. Once inside $|Q|< 1/\sqrt{3}$, the tails move slower than the Born distribution (which eventually catches up). So although the widened distribution may produce the most dramatic deviations from standard quantum mechanics, the deviations are short-lived and any measuring device would need to make its measurement before the tails recede.

In contrast, the narrowed distribution shows less dramatic behaviour. It recedes slowly to the back of the Born distribution, and then some is swept out, up and around the Born distribution (see the right-hand side of figure \ref{ground_sims}). The pointer stays roughly where one would expect it to from standard quantum mechanics. 

If one were to perform an ensemble of similar preparations and measurements, recording the position of the pointer in each, one would find the marginal distribution $\rho_{\text{mar}}(Y)$. The marginal distributions for $w=1/4,1$ and $4$ are shown at $T=4$ in figure \ref{marginals}. Any deviation that this distribution shows from the marginal Born distribution would of course be indicative of quantum nonequilibrium.

\subsubsection{Long-time/large $g$ measurement of vacuum modes}\label{long_time}
\begin{figure}
\input{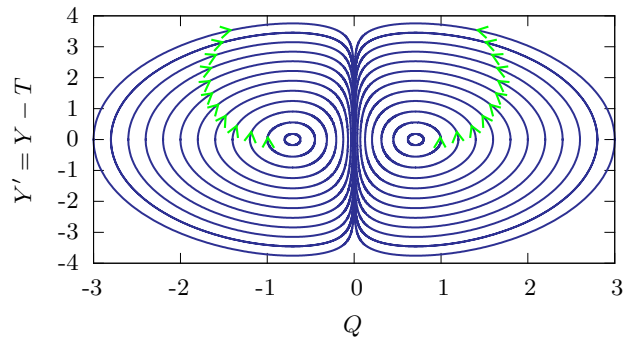}\\
\caption{\label{groundstate_trajectories}Above, a selection of the trajectories for the measurement of a vacuum mode (with normal ordering). The velocity field is time independent, resulting in periodic orbits around $(\pm\sqrt{1/2},0)$. Numerical simulations show that the pointer marginals converge to stationary nonequilibrium distributions characteristic of the initial nonequilibrium state (see figure \ref{long_time_marginals}).}
\end{figure}
Let us discuss a second measurement regime, which may be thought of as valid for large $T$ and/or (since $T=g  \omega t/(2\sigma)$) large $g$. 

\begin{figure}
\input{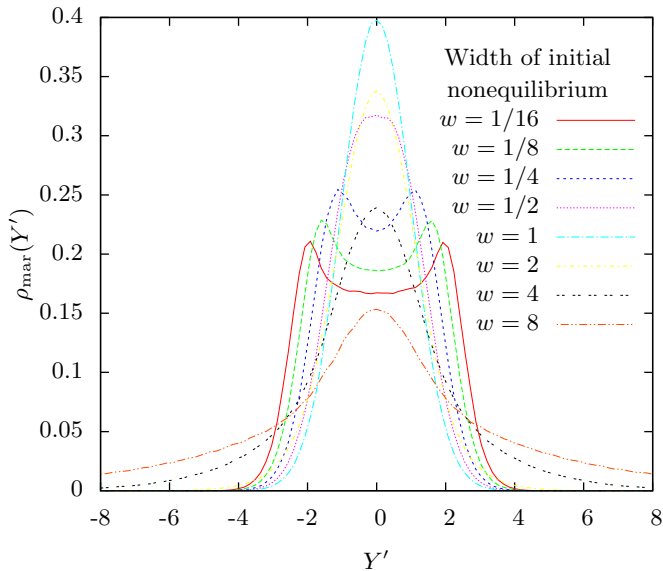}
\caption{\label{long_time_marginals}Characteristic stationary pointer marginals $\rho_{\text{mar}}(Y')$ for energy measurement of nonequilibrium vacuum modes in the large $T$ or large $g$ approximation. In this regime, initial nonequilibrium in the field mode will produce a corresponding stationary nonequilibrium for the pointer. Field modes with larger spread produce pointer marginals with larger spread. Field modes with smaller spread form pointer marginals with central depressions.}
\end{figure}
To aid analysis, we shall continue as if we had normal ordered the Hamiltonian \eqref{hamiltonian}. This, as mentioned in section \ref{sec:cont_eq}, is equivalent to switching to the `reference frame' of the Born distribution with $Y\rightarrow Y'= Y-T$. Under normal ordering the wave function and guidance equations become
\begin{align}
\Psi&=2^{-\frac14}\pi^{-\frac12}\exp\left(-\frac12 Q^2 -\frac14Y'^2 \right),\\
\partial_TQ&=-\frac13 QY', \label{gs_q_vel_y'}\\
\partial_TY'&=\frac{2}{3}Q^2-\frac{1}{3}.\label{gs_y_vel_y'}
\end{align}
The guidance equations are now time-independent and conserve a stationary Born distribution.
The trajectories are periodic.
A selection of the trajectories produced by equations \eqref{gs_q_vel_y'} and \eqref{gs_y_vel_y'} are shown in figure \ref{groundstate_trajectories}.
The trajectories do not pass the line $Q=0$, and so we cannot find relaxation to the Born distribution for any initial $\rho$ asymmetric in $Q$.

Our numerical simulations indicate that any nonequilibrium in the vacuum mode will, in the large $T$ or large $g$ limit, produce a corresponding stationary nonequilibrium in the pointer distribution. Furthermore, from this pointer distribution, numerical simulations could deduce the initial nonequilibrium in the vacuum mode. Our simulations show that this limit will be reached at $T\sim 120$ for $1/8<w<8$. 

Eight such stationary pointer marginals are displayed in figure \ref{long_time_marginals}. These are found under the measurement of nonequilibrium vacuum modes described by width parameters ranging from $w=1/16$ to $8$. Nonequilibrium modes that are wider than equilibrium make the spread in the pointer position correspondingly wider. In contrast, for the measurement of nonequilibrium vacuum modes that are narrower than equilibrium, the pointer marginal forms a central depression whilst staying in the same region. Measurements of the pointer over an ensemble would be enough to deduce the character of the initial nonequilibrium for each case.

\subsubsection{Measurement of a single particle state}
\begin{figure*}[t]
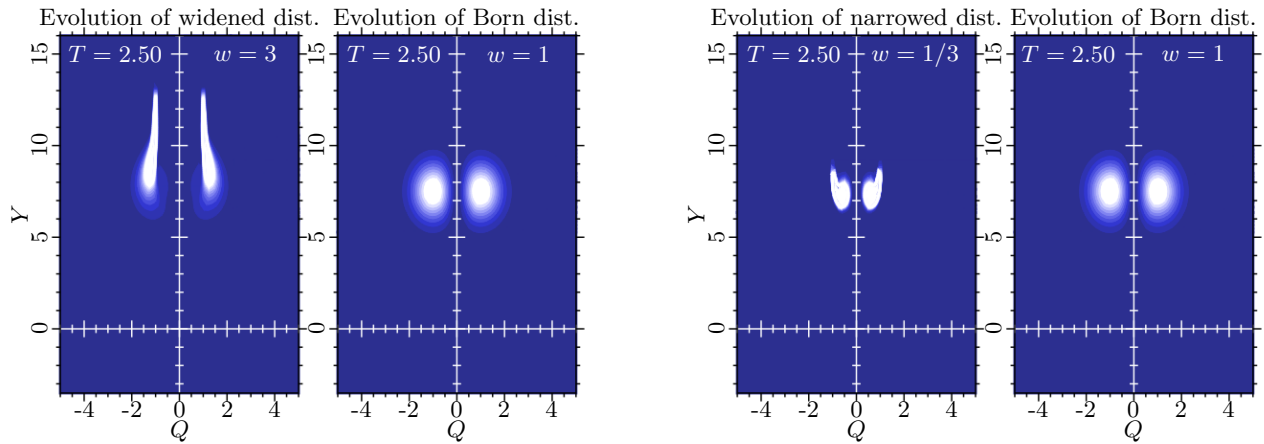

\input{excited_widened.eps_tex}
\input{excited_narrowed.eps_tex}
\caption{\label{excited_sims}The evolution of nonequilibria under energy measurement of single-particle states. On the left, a widened ($w=3$) nonequilibrium distribution; on the right, a narrowed ($w=1/3$) nonequilibrium distribution. The Born distribution, shown for comparison in each case, moves at a rescaled speed of $dY/dT=3$ (although individual de Broglie trajectories have variable speeds). Pointer marginal distributions for this process are shown in figure \ref{excited_marginal}.}
\end{figure*}
\begin{figure}
\input{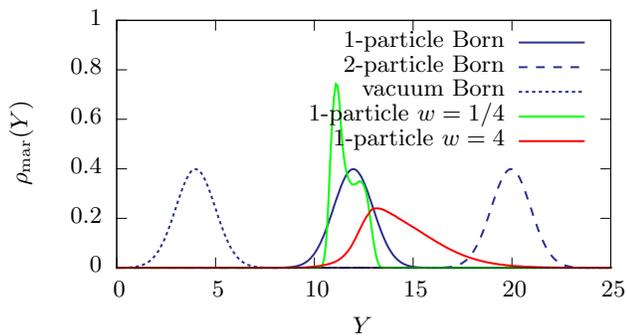}
\caption{\label{excited_marginal}Marginal pointer distributions $\rho_{\text{mar}}(Y)$ under energy measurement of one particle state nonequilibria at time $T=4$. (For comparison we also show the Born pointer marginals for the vacuum, 1-particle and 2-particle cases.) The widened nonequilibrium ($w=4$) shows a significant probability of detecting two excitations (or `particles') instead of one, and again there is a significant probability of finding the pointer around $Y= 16$ (a position disallowed by standard quantum mechanics for any initial superposition). As in the case of the vacuum mode measurement, the narrowed nonequilibrium ($w=1/4$) will be distinguished only by its internal structure. The tendency to form a double-bump in the pointer distribution is also seen in this case.}
\end{figure}
\begin{figure*}[t]
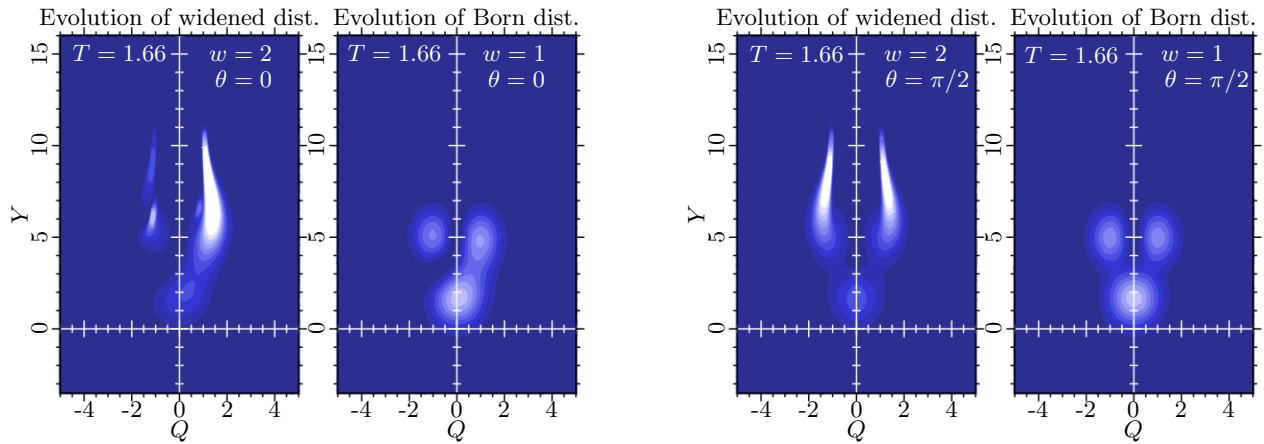

\input{super_th0.eps_tex}
\input{super_thpi2.eps_tex}
\caption{\label{super_sims}Evolution of joint distributions $\rho(Q,Y,T)$ under energy measurements of a nonequilibrium field mode in a superposition of a vacuum and a one-particle state with $c_0=e^{i\theta}/\sqrt{2}$ and $c_1=1/\sqrt{2}$ (Eq.\ \eqref{super1}). On the left we have taken $\theta=0$. On the right we have taken $\theta=\pi/2$. Both cases have widened distributions with $w=2$, and snapshots are taken at $T=1.66$. (For comparison, Born distributions are also shown in both cases.)}
\end{figure*}
\begin{figure}
\input{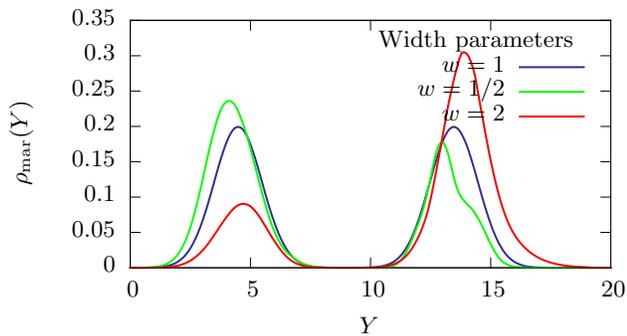}
\caption{\label{super_marginal}Marginal pointer distributions $\rho_{\text{mar}}(Y)$ for $c_0=c_1=1/\sqrt{2}$ and $w=1/2,1,2$, taken at $T=4.5$. Nonequilibrium is seen to cause anomalous spectra as observed by an experimenter. Similar results are obtained for other relative phases.}
\end{figure}
Under the energy measurement process, the effective wave function becomes Eq.\ \eqref{sp_measurement} and the trajectories satisfy the guidance equations \eqref{e1} and \eqref{e2}. The Born distribution evolves in the $Y$ direction at a rescaled velocity $dY/dT=3$. Since now the $Y$ velocity (Eq.\ \eqref{e2}) has terms proportional to $Q^2$ and $1/Q^2$, we might expect some increased pointer movement both for the widened and narrowed nonequilibrium cases.
In fact, our simulations show that a narrowed distribution yields relatively less pointer movement than the widened distribution (as we had for the case of the vacuum). Plots of the evolution of $\rho(Q,Y,T)$ are shown in figure \ref{excited_sims}, and marginal pointer distributions are shown in figure \ref{excited_marginal}. As in the case of the vacuum mode measurement, there is a significant probability of detecting an extra excitation or of finding the pointer in a position disallowed by standard quantum mechanics for any superposition being measured.

\subsubsection{Measurement of a superposition}
\begin{figure*}[t]
\input{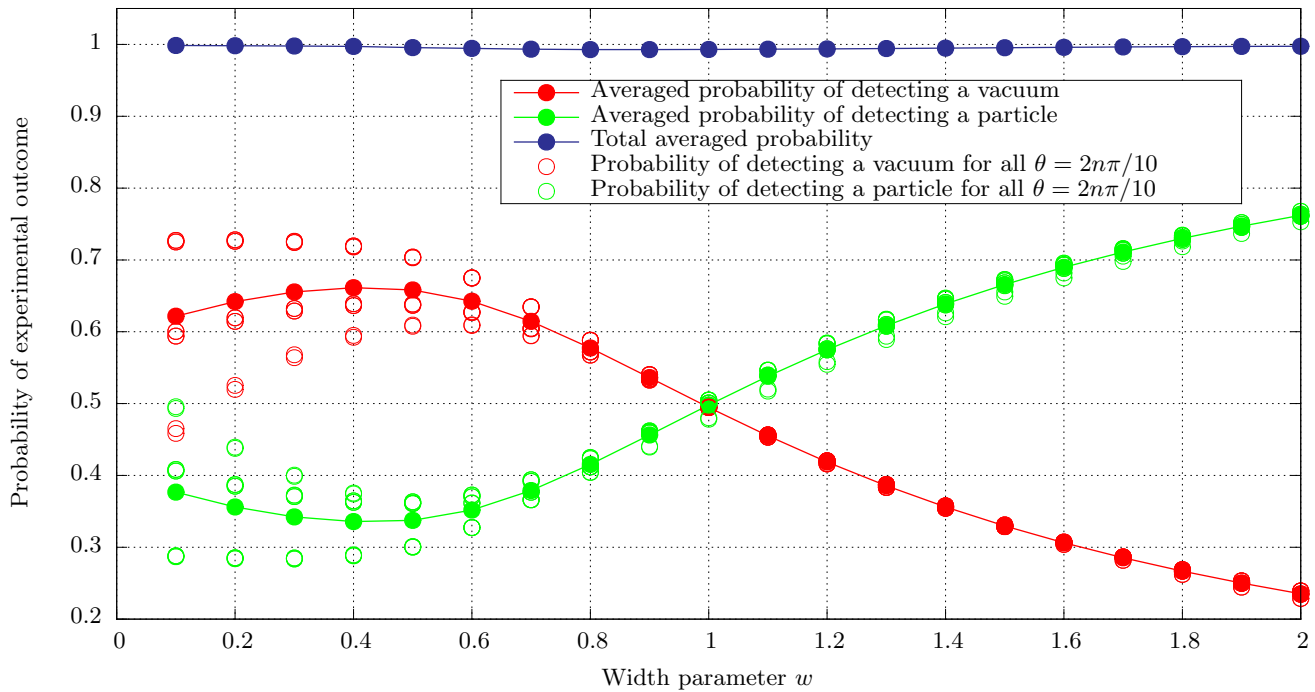}
\caption{\label{full_results}Ensemble probabilities of energy measurements for an equal superposition of particle and vacuum states as affected by quantum nonequilibrium of varying width $w$ (with results averaged over the relative phase $\theta$ in the superposition). As $|c_0|=|c_1|=1/\sqrt{2}$, there should be a 50\% probability of detecting a particle. However, widened nonequilibria give probabilities larger than 50\% for particle detection, while narrowed non-equilibria give probabilities less than 50\% for particle detection when averaged over $\theta$. (Hollow markers represent results for individual relative phases $\theta$, whilst solid markers represent averages over $\theta=2n\pi/10$, $n=1,2,\dots,10$. Dependence on the relative phase is seen to affect the outcomes only for $w\lesssim 1$.)}
\end{figure*}

Quantum nonequilibrium would in general cause anomalous results for the spectra of energy measurements. To illustrate this, we take the simple example of an equal superposition of vacuum and one-particle states. Quantum mechanically, an experimenter would observe a $50\%$ probability of detecting a particle. We take $c_0=e^{i\theta}/\sqrt{2}$ and $c_1=1/\sqrt{2}$, with the wave function and velocity fields specified in Eqs.\ (\ref{super1}-\ref{super3}). The dynamics of the measurement depends strongly on the initial relative phase $\theta$ of the superposition. This is seen in figure \ref{super_sims}, where we show the time evolution of joint distributions $\rho(Q,Y,T)$. Examples of the marginal pointer distributions produced in the energy measurement process are shown in figure \ref{super_marginal}. After about $T=3.5$, all marginal pointer distributions display two distinct areas of support, meaning that an experimenter would unambiguously obtain either $\frac12\omega$ or $\frac32\omega$ in each individual energy measurement, regardless of whether nonequilibrium is present or not. However, a widened nonequilibrium distribution would cause a larger probability of obtaining the outcome $\frac32\omega_\mathbf{k}$ (`detecting a particle'), while a narrowed nonequilibrium distribution would cause the opposite effect. Although the trajectories are strongly dependent on the initial phase, the marginal pointer distributions are only weakly dependent on this. 

In practice, one might not know the initial relative phase of the superposition. To make contact with what an experimenter might actually measure (albeit in the context of our simplified field-theoretical model), we have taken an average over $10$ phases: $\theta=2\pi n/10,\,n=0,1,\dots, 9$. We run each simulation up to time $T = 4.5$ and calculate the proportion of the distribution $\rho$ that lies beyond $Y=9.0$. This is the probability of observing an excitation, whilst the proportion of $\rho$ before $Y=9.0$ is the probability of observing the vacuum. (These numbers are clear from figure \ref{super_marginal}.) 
Figure \ref{full_results} illustrates the results of this averaging process for 20 separate width parameters $w$. We find a remarkable correlation. For example, for nonequilibrium close to the Born distribution, widening the distribution will proportionally increase the ensemble probability of `detecting a particle'. Clearly, nonequilibrium would generate an incorrect energy spectrum.

\section{Conclusion}
We have considered the possibility that our universe contains quantum nonequilibrium systems -- in effect a new form or phase of matter (including the vacuum) that violates the Born probability rule and which is theoretically possible in the de Broglie-Bohm formulation of quantum theory. While the practical likelihood of detecting such systems remains difficult to evaluate, we have argued that at least in principle they could exist today as relics from the very early universe. We have provided simple field-theoretical models illustrating the effects of quantum nonequilibrium in a particle-physics context. In particular, we have seen that quantum nonequilibrium would generate anomalous spectra for standard measurements of energy, as well as generating corrections to particle-physics processes generally.

The possibility of detecting relic nonequilibrium systems today depends on uncertain features of high-energy physics and cosmology. Dark matter, which is thought to make up approximately 25\% of the mass-energy of the universe, may consist of relic particles (such as gravitinos) that were created in the very early universe and which have propagated essentially freely ever since. (For reviews see, for example, refs. \cite{BHS05,EO13}.) As we have seen, such particles are plausible candidates for carriers of primordial quantum nonequilibrium and we expect that particle-physics processes involving them -- for example, decay or annihilation -- would display energetic anomalies.

On the experimental front, an especially promising development would be the detection of photons from dark matter decay or annihilation. These are expected to form a sharp spectral line, probably in the gamma-ray region. Recent interest has focussed on reports of a sharp line from the Galactic centre at $\sim130\ \mathrm{GeV}$ in data from the \textit{Fermi} Large Area Telescope (LAT) \cite{Betal12,W12}. While the line might be a dark matter signal, its significance (and even its existence)\ is controversial. The line could be caused by a number of scenarios involving dark matter annihilations \cite{BH12}. It might also be due to decaying dark matter \cite{ITW13}, for example the decay of relic gravitinos \cite{L13,Aetal14}. (In a supersymmetric extension of the Standard Model with violation of R-parity, the gravitino is unstable and can decay into a photon and a neutrino \cite{TY00}.) On the other hand, a recent analysis of the data by the Fermi-LAT team casts doubt on the interpretation of the line as a real dark matter signal \cite{FLAT13}.

Should dark matter consist (if only partially) of relic nonequilibrium systems, we may expect to find energetic anomalies for decay and annihilation processes. However, to distinguish these from more conventional effects would require more detailed modelling than we have provided here. There is also the question of whether the anomalies are likely to be large enough to observe in practice. These are matters for future work.

In principle, it would be of interest to test dark matter decay photons for possible deviations from the Born rule (perhaps via their polarisation probabilities \cite{AV04a}). We have seen that simple perturbative couplings will transfer nonequilibrium from one field to another, leading us to expect that in general a decaying nonequilibrium particle will transfer nonequilibrium to its decay products. Another open question, however, is the degree to which the nonequilibrium might be degraded during this process. In a realistic model of a particle decay we might expect some degree of relaxation. It would be useful to study this in pilot-wave models of specific decay processes.

As a general point of principle, one might also be concerned that in the
scenario discussed in this paper the probability distribution for delocalised
field modes in the early universe -- where the probability distribution is
presumably defined for a theoretical ensemble -- appears to have measurable
implications for decay particles in our one universe. How can this be? A
similar point arises in the standard account of how the power spectrum for
primordial perturbations has measurable implications for our one CMB sky. In
inflationary theory, the probability distribution for a single mode
$\phi_{\mathbf{k}}$ of the inflaton field does have measurable implications in
our single universe. As we discussed in Section \ref{IIA}, the variance
$\left\langle |\phi_{\mathbf{k}}|^{2}\right\rangle $ of the primordial
inflaton distribution appears in the power spectrum $\mathcal{P}_{\mathcal{R}%
}(k)\propto\left\langle |\phi_{\mathbf{k}}|^{2}\right\rangle $ for primordial
curvature perturbations $\mathcal{R}_{\mathbf{k}}\propto\phi_{\mathbf{k}}$ at
wave number $k$. The power spectrum $\mathcal{P}_{\mathcal{R}}(k)$ in turn
appears in the angular power spectrum $C_{l}$ (equation \eqref{Cl2}), which may be
accurately measured for our single CMB sky provided $l$ is not too small. In
the standard analysis it is assumed that the underlying `theoretical ensemble'
of universes is statistically isotropic, which implies that the ensemble
variance $C_{l}\equiv\left\langle \left\vert a_{lm}\right\vert ^{2}%
\right\rangle $ is independent of $m$ -- where $a_{lm}$ are the harmonic
coefficients for the observed temperature anisotropy. We then in effect have
$2l+1$ measured quantities $a_{lm}$ with the same theoretical variance.
Provided $l$ is sufficiently large, one can perform meaningful statistical
tests for our single CMB sky and compare with theoretical predictions for
$C_{l}$. Statistical homogeneity also plays a role in relating the $C_{l}$'s
for a single sky to the power spectrum $\mathcal{P}_{\mathcal{R}}(k)$ for the
theoretical ensemble \cite{AV10,Muk05}. To understand how the
theoretical ensemble probability has measurable implications in a single
universe, it is common to speak of the CMB sky as divided up into patches --
thereby providing an effective ensemble in one sky. This works if $l$ is
sufficiently large, so that the patches are sufficiently small in angular
scale and therefore sufficiently numerous. Similar reasoning applies to
particles (or field excitations) generated by inflaton decay. In this context
it is important to note that realistic particle states, as observed for
example in the laboratory, are represented by field modes defined with respect
to finite spatial volumes $V$. Almost all of the particles in our universe
were created by inflaton decay, and in practice their states are in effect
defined with respect to finite spatial regions. By measuring particle
excitations in different spatial regions, it is possible to gather statistics
for outcomes of (for example)\ energy measurements. (One might also consider a
time ensemble in one region, but a space ensemble seems more relevant in the
case of relic decay particles.) The resulting statistical distribution of
outcomes for the decay particles will depend on the original probability
distribution for the decaying inflaton field -- just as the statistics for
patches of the CMB sky depend on the probability distribution for the inflaton
during the inflationary era. A full account would require an analysis of
inflaton decay more precise than is currently available. In particular, one
would like to understand how this process yields particle states that are
confined to finite spatial regions. It is generally understood that the decay
products form as excitations of sub-Hubble modes, with wave functions confined
to sub-Hubble distances. Depending on the details, this can correspond to
relatively small spatial distances today.
Of course,
particle wave packets will also spread out since their creation, but still we
may expect them to occupy finite spatial regions. Further elaboration of this
point lies outside the scope of this paper.

Even if there exist localised sources or spatial regions containing particles in a state of quantum nonequilibrium, it might be difficult in practice to locate those regions. In particular, if a given detector registers particles belonging to different regions without distinguishing between them, then it is possible that even if nonequilibrium is present in the individual regions it will not be visible in the data because of averaging effects. How one might guard against this in practice remains to be studied.

Finally, we have seen that the likelihood of nonequilibrium surviving until today for relic particles depends on the fact that a nonequilibrium residue can exist in the long-time limit for systems containing a small number of superposed energy states \cite{ACV14}. While this may certainly occur in principle, its detailed implementation for realistic scenarios requires further study. On the other hand, no such question arises in our scenario for relic nonequilibrium vacuum modes, since the simplicity of vacuum wave functionals guarantees that further relaxation will not occur at late times. Long-wavelength vacuum modes may be carriers of primordial quantum nonequilibrium, untouched by the violent astrophysical history that (according to our hypotheses) long ago drove the matter we see to the quantum equilibrium state that we observe today. It remains to be seen if, in realistic scenarios, the effects on particle-physics processes taking place in a nonequilibrium vacuum could be large enough to be detectable.

\begin{acknowledgments}
This research was funded jointly by the John Templeton Foundation and Clemson University.
\end{acknowledgments}
\appendix*
\section{Numerical methodology}\label{code}
Most studies of relaxation in de Broglie-Bohm theory have used the back-tracking method of ref.\ \cite{VW05} (see for instance \cite{VW05,TRV12,SC12,CV13}).
This method uses the fact that the ratio $f=\rho(\mathbf{x},t)/|\psi(\mathbf{x},t)|^2$ is conserved along trajectories. A uniform grid of final positions is evolved backwards from the final time $t_f$ to the initial time $t_i$. The final distribution is constructed from the conserved function $f$. Although this method has been successful in producing accurate results, it has the disadvantage that backtracking to $t_i$ must be carried out for each desired final time $t_f$.

We have instead chosen to integrate the continuity equation \eqref{ant_cont} directly using a finite-volume method. The method used is a variant of the corner transport upwind method detailed in sections 20.5 and 20.6 of \cite{leve02}, modified so as to apply to the conservative form of the advection equation. This algorithm has the advantage that different `high resolution limiters' may be switched off and on with ease, so that one may compare results. (We use a monotonised central (MC) limiter throughout.) The main disadvantage of this approach is a consequence of the velocity field \eqref{guidance_a} and \eqref{guidance_b} diverging at nodes (where $|\psi|\rightarrow 0$). Since such an algorithm is required to satisfy a Courant-Friedrichs-Lewy condition to maintain stability, without velocity field smoothing the algorithm is inherently unstable.
We have found that a simple way to implement a smoothing is to impose a maximum absolute value on the velocities. The maximum is taken throughout to be $1/10^{\text{th}}$ of the ratio of grid spacing to time step.

We have found that the finite-volume method is less efficient than the backtracking method over larger time scales. 
In fact, the long-time simulations shown in figure \ref{long_time_marginals} were produced using a fifth-order Runge-Kutta algorithm to evolve trajectories directly.
 However for short time scales -- the prime focus of this work -- the finite-volume method is a useful tool.

%

\end{document}